%% file: arXiv_PH_of_DN_article.tex
\documentclass[smallextended]{svjour3}       %
\smartqed  %
\usepackage{newtxtext,newtxmath}  %
\usepackage{ulem}

\newcommand{\revisionRemove}[1]{\sout{#1}}
\newcommand{\revision}[1]{{\color{purple} #1}}
\renewcommand{\revision}[1]{{\color{black} #1}}
\renewcommand{\revisionRemove}[1]{}
\usepackage{my}
\hypersetup{
    colorlinks=true,
    linkcolor= violet,
    urlcolor= violet,
    citecolor = violet
}
\begin{document}

\title{Temporal Network Analysis Using Zigzag Persistence}
\titlerunning{Temporal Network Analysis Using Zigzag Persistence}

\author{Audun D. Myers         \and
        David E. Muñoz 	  \and
        Firas A. Khasawneh$^*$ 	  \and
        Elizabeth Munch 	
}

\institute{A. Myers \at
              Michigan State University, College of Engineering \\
              \email{myersau3@msu.edu}           %
           \and
           D. Muñoz \at
              Michigan State University, College of Engineering
           \and
           F. Khasawneh ($^*$corresponding author) \at
              Michigan State University, College of Engineering \\
              \email{khasawn3@egr.msu.edu}           %
           \and
           E. Munch \at
              Michigan State University, College of Engineering 
}

\date{Preprint as of \today}

\maketitle

\begin{abstract}
\input{sections/sec-abstract}

\end{abstract}

\input{./sections/sec-intro}

\input{./sections/sec-background}

\input{./sections/sec-method}

\input{./sections/sec-results}

\input{./sections/sec-conclusion}

\begin{acknowledgements}
\input{./sections/sec-acknowledgements}
\end{acknowledgements}

\section*{Conflict of interest}
On behalf of all authors, the corresponding author states that there is no conflict of interest.

\bibliographystyle{unsrt}
\bibliography{dyn_net_zz_bib}
\appendix
\input{./sections/sec-appendix}
\end{document}

%% file: sections/sec-abstract.tex
This work presents a framework for studying temporal networks using zigzag persistence, a tool from the field of Topological Data Analysis (TDA). 
The resulting approach is general and applicable to a wide variety of time-varying graphs. For example, these graphs may correspond to a system modeled as a network with edges whose weights are functions of time, or they may represent a time series of a complex dynamical system. 
We use simplicial complexes to represent snapshots of the temporal networks that can then be analyzed using zigzag persistence. 
We show two applications of our method to dynamic networks: an analysis of commuting trends on multiple temporal scales, e.g., daily and weekly, in the Great Britain transportation network, and the detection of periodic/chaotic transitions due to intermittency in dynamical systems represented by temporal ordinal partition networks. 
Our findings show that the resulting zero- and one-dimensional zigzag persistence diagrams can detect changes in the networks' shapes that are missed by traditional connectivity and centrality graph statistics.

%% file: sections/sec-intro.tex
\section{Introduction} \label{sec:intro}

Network data, that is, the encoding of connections between objects, is a natural form of information representation in many fields \cite{Porter2020}.
While the analysis of static networks or graphs is already a broad field of research in itself, there is often much information ignored. 
In particular, we are interested in the case of \textit{temporal networks} \cite{Holme2013,Holme2015}; that is, the case of a dynamical system represented by a network evolving over time. 
These networks can arise in many different cases, such as 
social networks~\cite{Skyrms2000}, 
disease spread dynamics~\cite{Husein2019}, 
manufacturer-supplier networks~\cite{Xu2019b}, 
power grid network~\cite{Schaefer2018}, and
transportation networks~\cite{DavidBoyce2012}. 
Many important characteristics of a dynamical network can be extracted from the data. These include source and rate of disease spread as well as predictions on future infections~\cite{Enright2018}, weak branches in supply chains and possible failures~\cite{Nuss2016, Xu2019b}, changes in infrastructure to avoid cascade failures in power grids~\cite{Soltan2014, Schaefer2018}, transportation network optimal routing (finding an optimal minimum time route between)~\cite{Bast2015}, fault analysis (detecting transportation disruptions) in transportation networks~\cite{Sugishita2020}, and flow pattern analysis (visualization)~\cite{Hackl2019}.

Temporal graph data is commonly represented using attributed information on the edges for the time intervals or instances in which the edges are active~\cite{Chen2016a,Huang2015}. 
Using this attributed information, we can represent the graph in several ways including edge labeling, snapshots, and static graph representations~\cite{Wang2019}. 
In this work we will first represent the data in the standard attributed (labeled) temporal graph structure and then use the graph snapshots approach, where our input is a sequence of static graphs $G_0, G_1, \ldots, G_n$.

The available tools for analyzing such data is often inspired by the tools from the static network community; for instance, 
centrality or flow measures~\cite{Borgatti2005}; temporal clustering for event detection~\cite{Crawford2018, You2021, Moriano2019}; and connectedness~\cite{Kempe2002}. 
However, these tools do not account for higher-dimensional structures (e.g., loops as a one-dimensional structure). 
It may be important to account for evolving higher-dimensional structures in temporal networks to understand the changing structure better. For example, a highly connected network may only have one connected component with no clear clusters, but the number of loops within the network may detect the change.

For this reason, we introduce a method for incorporating ideas from Topological Data Analysis (TDA) \cite{Dey2021,Munch2017} to encode more complex structure than can be seen in the standard graph tools. 
The mainstay of TDA is persistent homology, colloquially referred to as persistence, which  encodes structure by analyzing the changing shape of a simplicial complex (a higher dimensional generalization of a network) over a filtration (a nested sequence of subcomplexes $K_1 \subseteq K_2 \subseteq \cdots \subseteq K_n$).
This shape is measured via homology, a vector space encoding information about topological structure of the space. 
Different dimensions of homology measure different things: 0-dimensional homology encodes information about connected components; 1-dimensional homology encodes loops; 2-dimensional homology encodes voids. 
How these structures change over the filtration (e.g.~when a loop appears and then subsequently fills in) can be stored as a persistence diagram; that is, a collection of points in the upper half plane where a point at $(i,j)$ means that a structure appeared at $K_i$ but was lost at $K_j$.

However, a major limitation of standard persistence for the temporal network input data is the requirement that inclusions only go one way. 
Over the evolution parameter, our temporal graphs might add or remove edges, and we would like to build a system that can account for this. 
Thus, to study the evolving higher dimensional structures within a temporal network, we will leverage zigzag persistence~\cite{Carlsson2010}, which allows for insertions and deletions of simplices at every step.
The same mathematical theorems that make it possible to represent the information of standard persistence in a persistence diagram or barcode can be used in the case of a zigzag filtration. 
Zigzag persistence tracks the formation and disappearance of homological structures through a persistence diagram as a two-dimensional summary diagram. 

The tools we employ will treat the graph input itself as a topological structure by using it to extract distance information between the vertices. 
For the work discussed here, we will assume the input networks are unweighted graphs and thus use the shortest path distance between vertices. 
However, this is by no means the only way to incorporate graph data for persistent homology; see \cite{Aktas2019} for a survey. 
Further, because we generally treat our networks as a metric space, our intuition is that the tools developed here are most applicable to data where vertices have a geometric component (such as networks where nodes come from geo-spatial data) rather than only combinatorial data (such as human proximity networks, even if this has to do with physical proximity), but we look forward to being proven wrong.

We thus apply our methods to two sample data sets. 
The first is transportation data from Great Britain \cite{Gallotti2015}, where we can use the zigzag persistent homology for understanding higher order structures in the system, and see behavior such as daily periodicity represented automatically in the diagram output. 
The second data comes from input data from time series analysis. 
We have previously used zigzag persistence to detect Hopf bifurcations  by taking as input a point cloud approximation of the underlying attractor to the time series \cite{Tymochko2020}. 
However, the computational limitations of zigzag persistence on large point clouds makes its use in this setting unwieldy. 
Thus, we combine these ideas with recently available network representations of time series \cite{McCullough2015} to 
instead encode the structure of the time series in a graph.
Previous work shows that measuring this structure using standard persistent homology can be used to differentiate between behaviors in the underlying dynamical system \cite{Myers2019,Myers2022}.

\subsection{Organization}

We will start in Section~\ref{sec:background} with an introductory background on persistent homology and zigzag persistence.
Next, in Section~\ref{sec:method},  we overview the general pipeline for applying zigzag persistence to temporal graph data. We couple this explanation with a demonstrative toy example. 
In Section~\ref{sec:results} we  introduce the two systems we will study. 
The first is a dataset collected over a week of the Great Britain transportation system~\cite{Gallotti2015}. 
The second is an intermittent Lorenz system simulation, where we generate a temporal network through complex networks of sliding windows. %
Then we apply zigzag persistence to our two examples and show how the resulting persistence diagrams help visualize the underlying dynamics in comparison to standard temporal network analysis techniques.

%% file: sections/sec-background.tex
\section{Background} \label{sec:background}

\subsection{Persistent Homology}
Persistent homology, the flagship tool from the field of Topological Data Analysis (TDA), is used to measure the shape of a dataset at multiple dimensions. For example, it can measure connected components (dimension zero), loops (dimension one), voids (dimension two), and higher dimensional analogues. 
Persistent homology measures these shapes using a parameterized filtration to detect when the structures are born (appear) and die (disappear). 
We give the basic ideas in this section and direct the interested reader to more complete introductions to 
standard homology \cite{Hatcher,Munkres2}
and persistent homology \cite{Dey2021,Oudot2015,Munch2017}.

The required input for persistence is a filtration of a simplicial complex $K$. 
Specifically,  a filtration $\{K_{a_i}\}_i$ is a parameterized sequence of simplicial complexes which are nested; i.e.
\begin{equation}
K_{\alpha_0} \subseteq K_{\alpha_1} \subseteq K_{\alpha_2} \subseteq \ldots \subseteq K_{\alpha_n}.
\label{eq:nested_complexes}
\end{equation}
Under this notation, there are $n+1$ simplicial complexes; and it is often the case that $K_{\alpha_0}$ is either the empty complex or the complex consisting of only the vertex set. 
We can then calculate the homology of dimension $p$ for each complex, $H_p(K_{a_i})$, which is a vector space representing the $p$-dimensional structure of the space such as loops, voids, etc.
Surprisingly, there is further information to be used, namely that the inclusions on the simplicial complexes induce linear maps on the vector spaces resulting in a construction called a persistence module: 
\begin{equation}
H_p(K_{\alpha_0}) \to H_p(K_{\alpha_1}) \to H_p(K_{\alpha_2}) \to \ldots \to H_p(K_{\alpha_n}).
\label{eq:filtration}
\end{equation}

The appearance and disappearance of classes in this object can be tracked, resulting in a representation of the information known as a persistence diagram. 
For each class which appears at $K_{b_i}$ and disappears at $K_{d_i}$, we draw a point in the plane at $(b_i,d_i)$. 
Taken together, the collection of points (also called persistence pairs), all of which are above the diagonal $\Delta= \{ (x,y) \mid x = y\}$, is called a persistence diagram.

One common approach to obtain this setup is in the case where the input data is a finite metric space $(M,d)$; i.e.~a finite set $M$ with distances given by $d(m_1,m_2)$. 
For example, $M$ might be a finite point cloud $\chi \subset \R^k$ with $d$ given by Euclidean distance. 
In the case of a graph $G$ as input data, we can have $M=V$ the vertex set, and $d(u,v)$ as the number of edges in the shortest path between vertex $u$ and vertex $v$. 

From this metric space data, we fix a value $\alpha \geq 0$ and construct the Vietoris-Rips (VR) complex, denoted $R_\alpha(M)$. 
In the VR construction, we have a vertex set $M$ and a simplex $\sigma \subseteq M$ is included in the abstract simplicial complex $R_\alpha(M)$ whenever $d(v,w) \leq \alpha$ for all $v,w \in \sigma$. 
Then by definition, $R_\alpha(\chi) \subseteq R_{\alpha'}(\chi)$ whenever $\alpha \leq \alpha'$. 
In this particular context, the axes of the persistence diagram correspond to distances between points.
So, for example, points in the diagram that are far from the diagonal of the 1-dimensional persistence diagram represent large loop structures in the input data. 

\begin{figure}
    \centering
    \includegraphics[width = \textwidth]{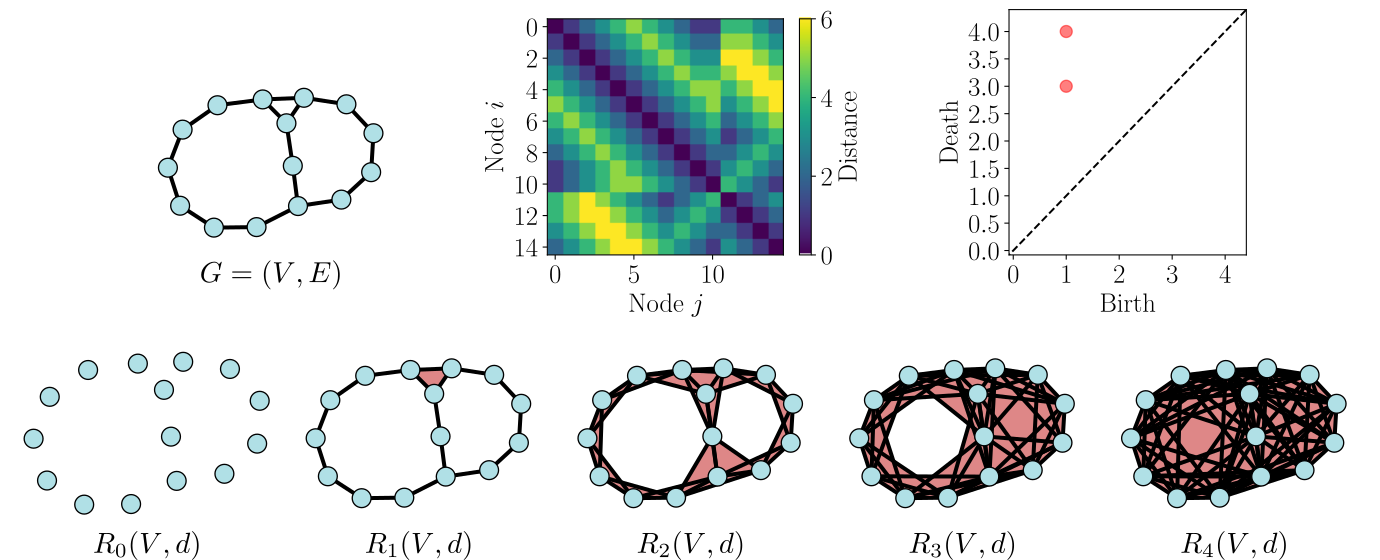}
    \caption{An example of a Rips filtration where the input is the shortest path distance on an example graph $G$. The persistence diagram in the top right is for standard persistence rather than zigzag as will be discussed later in the paper. }
    \label{fig:bigGraphPersistence}
\end{figure}
See Fig.~\ref{fig:bigGraphPersistence} for an example of the filtration in this setting. 
The graph $G$ is shown at the top left, and the distance defined on the nodes is given in the matrix at the top middle. 
The bottom row shows the Rips complexes for different choices of $\alpha$ parameter. 
In particular, note that $R_1(V,d)$ has the graph $G$ as the 1-skeleton, but includes an additional triangle not present in the graph. 
The top right shows the 1-dimensional persistence diagram. 
Note that the two large loops in the graph are encoded as points in the diagram; while the small triangle is immediately filled in and thus not represented.

\subsection{Zigzag Persistence}

A limitation of the standard setup of persistent homology is that it requires each simplicial complex to be a subset of the next, as shown in Eq.~\eqref{eq:nested_complexes}.
This means that at each step, we are only allowed to add new simplices to the previous complex to build this filtration. 
However, temporal graphs have no such behavior.
Thus, this issue can be alleviated through zigzag persistence~\cite{Carlsson2010, Carlsson2009a}, which allows for inclusions which can go either way at each step. 
This is often written as 
\begin{equation}
K_{\alpha_0} \leftrightarrow K_{\alpha_1} \leftrightarrow K_{\alpha_2} \leftrightarrow \ldots \leftrightarrow K_{\alpha_n},
\label{eq:bi_directional_zigzag_complexes}
\end{equation}
where $\leftrightarrow$ denotes one of the two inclusions $\hookrightarrow$ and $\hookleftarrow$. 

A common special case of this definition is where the left and right inclusions alternate, which can arise by taking a sequence of simplicial complexes, and interleaving them with either unions or intersections of the adjacent complexes.
Focusing on the case of the union, denote $K_{i,i+1} = K_i \cup K_{i+1}$ to get the zigzag filtration
\begin{equation*}
K_{\alpha_0} \hookrightarrow K_{\alpha_0,\alpha_1} 
\hookleftarrow K_{\alpha_1} 
\hookrightarrow K_{\alpha_1,\alpha_2} 
\hookleftarrow K_{\alpha_2} 
\hookrightarrow \ldots 
\hookleftarrow K_{\alpha_{n-1}}  
\hookrightarrow K_{\alpha_{n-1},\alpha_{n}} 
\hookleftarrow K_{\alpha_n}.
\end{equation*}

The same algebra that makes it possible for standard persistence to be represented in a diagram allows for computation of when homology features are born and die based on the zigzag persistence, however one must take care as some of the intuition from standard persistence is lost.
We can again track this with a persistence diagram consisting of persistence pairs $(b_i, d_i)$. 
In the case of a class appearing or disappearing at the union complex $K_{\alpha_{i},\alpha_{i+1}}$, we draw the index at the average $(\alpha_i+\alpha_{i+1})/2$.
If a topological feature persists through the last simplicial complex we set its death as the end time of the last window or index $n+0.5$. 

\begin{figure}%
    \centering
    \includegraphics[width=0.83\textwidth]{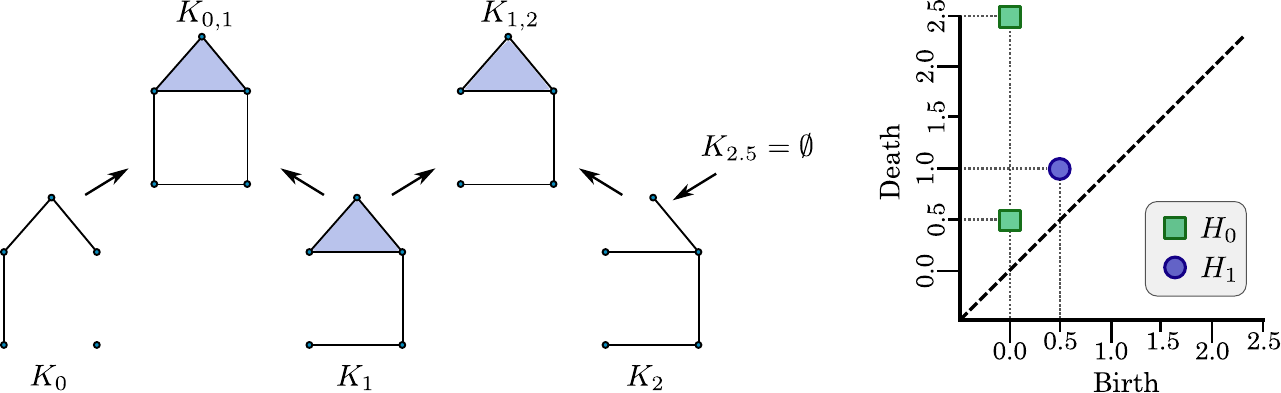}
    \caption{
    Example application of zigzag persistence to study changing topology of simplicial complex sequence. This example shows the sequence of simplicial complexes with intermediate unions and corresponding time stamps on the left and on the right the resulting zigzag persistence diagram on the right for dimension $0$ and $1$ as $H_0$ and $H_1$, respectively.
    }
    \label{fig:simple_zigzag_example}
\end{figure}
To demonstrate how zigzag persistence tracks the changing topology in a sequence of simplicial complexes we will use a simple example shown in Fig.~\ref{fig:simple_zigzag_example}. 
The sequence of simplicial complexes are shown as $[K_{0}, K_{1}, K_{2}]$, with unions given by 
$K_{0,1}$ and $K_{1,2}$.
The persistence diagram then tracks where topological features of various dimensions of $H_p$ (dimension 0 and 1 for this example) form and disappear. 
For example, for $H_0$ there are two components in $K_{0}$. 
At the next simplicial complex $K_{0,1}$ the two $0$-dimensional features combine signifying one of their deaths which is tracked in the persistence diagram as the persistence pair $(0,0.5)$ since 0.5 is the average of $0$ and $1$. 
The component that persists remains throughout all of the simplicial complexes. 
Therefore, we set its death as the last time stamp plus 0.5 and record the persistence pair as $(0, 2.5)$. 
On the other hand, there is a single loop (a 1-dimensional feature) which is shown only at $K_{0,1}$. 
For technical reasons, this is drawn as a point with birth time 0.5 (the average of $0$ and $1$) and death time 1 since it dies entering $K_{1}$. 

\begin{figure}%
    \centering
    \begin{minipage}{0.21\textwidth}
         \centering
         \includegraphics[width=\textwidth]{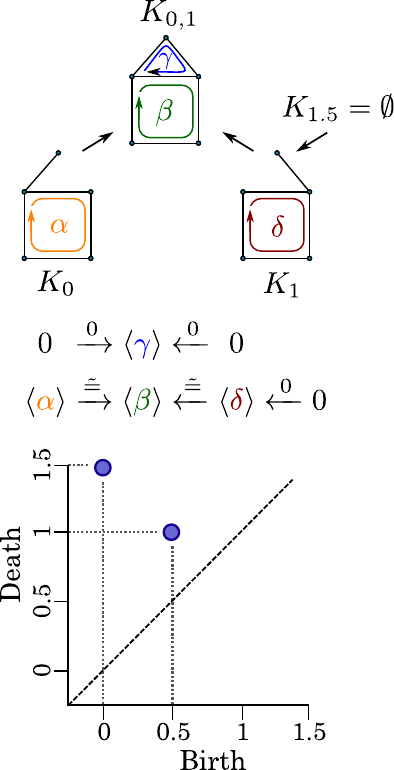} \\
         (a) Case 1
    \end{minipage}
    \hfill
    \begin{minipage}{0.21\textwidth}
         \centering
         \includegraphics[width=\textwidth]{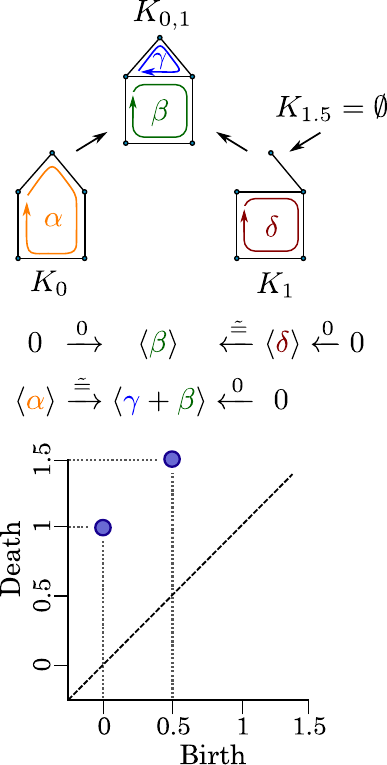} \\
         (b) Case 2
    \end{minipage}
    \hfill
    \begin{minipage}{0.21\textwidth}
         \centering
         \includegraphics[width=\textwidth]{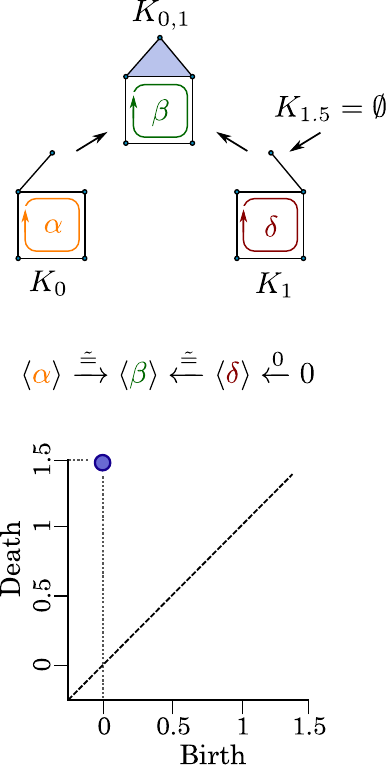} \\
         (c) Case 3
    \end{minipage}
    \hfill
    \begin{minipage}{0.21\textwidth}
         \centering
         \includegraphics[width=\textwidth]{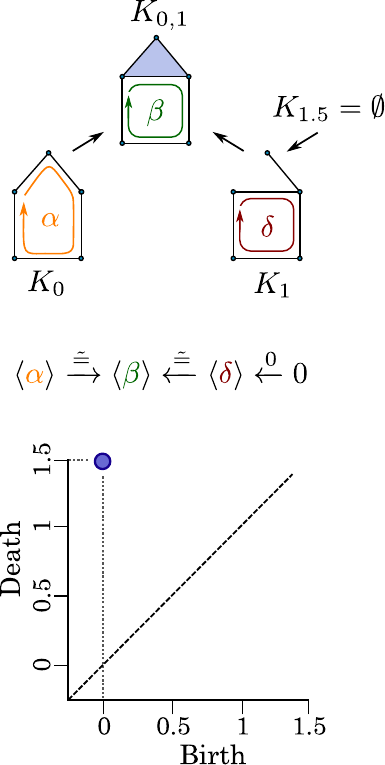} \\
         (d) Case 4
    \end{minipage}
    \caption{Several examples showing how the zigzag persistence diagram can be affected by the existence of small loops. 
    The first and third; and the second and fourth examples have the same filtration with the addition of a single triangle in the later case. 
    In each case, the generators of homology are shown with the interval decomposition drawn below. The resulting persistence diagrams show that the third and fourth examples show a single bar representing the persisting loop. 
    However in the first two intervals are quite different, where the long lived bar is split into two pieces in the second example.
    }
    \label{fig:HouseExample}
\end{figure}
It should be noted that sometimes zigzag persistence results are counter-intuitive to those used to standard persistence interpretation.  
For example, consider the example of Fig.~\ref{fig:HouseExample}.
The first two filtrations each appear to have a circular structure which lasts through the duration of the filtration, however, the zigzag persistence diagram sees the first case as a class living throughout, while the second breaks this class into two individual bars. 
In this case at least, the issue can be mitigated by including a triangle at the top of the house filling in the short cycle, resulting in both filtrations having only a single point in the persistence diagram representing the loop in the house. 

Secondly, we also note that the axes in the zigzag diagram correspond to indexing in the zigzag diagram.  
This means that a point far from the diagonal means there is a loop structure that is present for a large portion of the index set, rather than being a measurement of size of that same loop.

%% file: sections/sec-method.tex
\section{Method} \label{sec:method}
\begin{figure}
    \centering
    \includegraphics[width=0.97\textwidth]{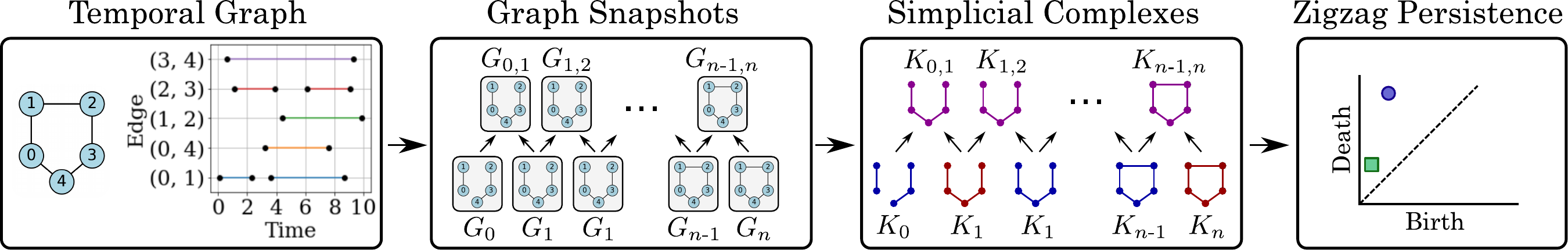}
    \caption{Pipeline for applying zigzag persistence to temporal networks. Begin with an unweighted and undirected \textbf{temporal graph} where each edge is on at a point or interval of time. Create \textbf{graph snapshots} using a sliding window interval over the time domain. Create a sequence of \textbf{simplicial complexes} from the graphs and apply \textbf{zigzag persistence} to the union zigzag simplicial complexes.}
    \label{fig:pipeline}
\end{figure}

To apply zigzag persistence for studying temporal graphs, we use the pipeline shown in Fig.~\ref{fig:pipeline} which we describe more specifically here.
A temporal graph is a graph structure that incorporates time information on when edges and/or nodes are present in the graph. 
We will only be using the case of temporal information attributed to the edges in this work and assume nodes are included as part of an edge-induced subgraph.
Thus, our starting data is a graph $G = (V,E)$ where each edge $e$ has a collection of closed intervals $\mathcal{I}_e = \{I_{e,1}, \cdots, I_{e,k_e} \}$ associated to it for when that edge is active. 
Fitting with the previous section, we fix a collection of times $t_0\leq t_1 \leq \cdots \leq t_n$ and a choice of window size $w$.
Then $G_{i} = (V_i,E_i)$ is the graph induced by edges present within a window $[t_i-w/2,t_i+w/2]$. 
Please note that to ease notation, we are using subscripts $i$ for the graphs, but the graphs should be thought of as encoding information for the interval centered at $t_i$ and thus when computing persistence, the birth and death times are associated to the $t_i$'s rather than the $i$'s. 
Finally, we define $G_{i,i+1} = G_{i} \cup G_{{i+1}}$ to be the union of the two adjacent graphs.

While we can construct a zigzag filtration of graphs 
\begin{equation*}
G_{0} 
\hookrightarrow G_{0,1} 
\hookleftarrow G_{1} 
\hookrightarrow G_{1,2} 
\hookleftarrow G_{2} 
\hookrightarrow \ldots 
\hookleftarrow G_{{n-1}}  
\hookrightarrow G_{{n-1},{n}} 
\hookleftarrow G_{n}
\end{equation*}
this is not actually the zigzag we will be using since, as noted earlier, it is not flexible enough to find some of the loop structures needed in the later analysis. 
So, for any graph $G_i$ with vertex set $V_i \subseteq V$, we let $d_i$ be the shortest path distance on that graph. 
That is, $d_i:V_i \times V_i \to \R$ where $d_i(u,v)$ is the number of edges needed to get from vertex $u$ to vertex $v$. 
Fixing an $r\geq 0$, let $K_{i}^r:=R_r(V_i, d_i)$ be the Rips complex constructed from this distance information. 
First, we note that $K_i^0$ is the simplicial complex with only the vertex set $V_i$. 
When $r=1$, $K_i^1$ is the clique complex of the original graph $G_i$; where the 1-simplices are the same as that of the graph, but higher dimensional simplices are filled in when available. 
Then for higher $r$, we have more and more edges added to the original graph, so this construction is more complicated than simply treating the graph itself as a simplicial complex.

Assuming we have similar notation for the union graph information ($d_{i,{i+1}}$, $K_{i,{i+1}}^r$, etc), we form a zigzag filtration by replacing $G_i$ with its Vietoris-Rips complex,
\begin{equation*}
 K_{0}^r %
\hookrightarrow K_{0,1}^r %
\hookleftarrow K_{1}^r %
\hookrightarrow K_{1,2}^r %
\hookleftarrow K_{2}^r %
\hookrightarrow \ldots 
\hookleftarrow K_{{n-1}}^r %
\hookrightarrow K_{{n-1},{n}}^r %
\hookleftarrow K_{n}^r. %
\end{equation*}
Finally, we compute the $p$-dimensional homology at each step and from there compute the zigzag persistence diagram. 
Our code uses the \texttt{Dionysus2} package \cite{Dionysus2} for this last step.
Again, be aware that the indices on the zigzag persistence points correspond to the $t_i$ value associated to the given complex.

\subsection{Example} 
\label{ssec:example}

\begin{figure}
     \centering
     \begin{minipage}{0.64\textwidth}
         \centering
         \includegraphics[width=\textwidth]{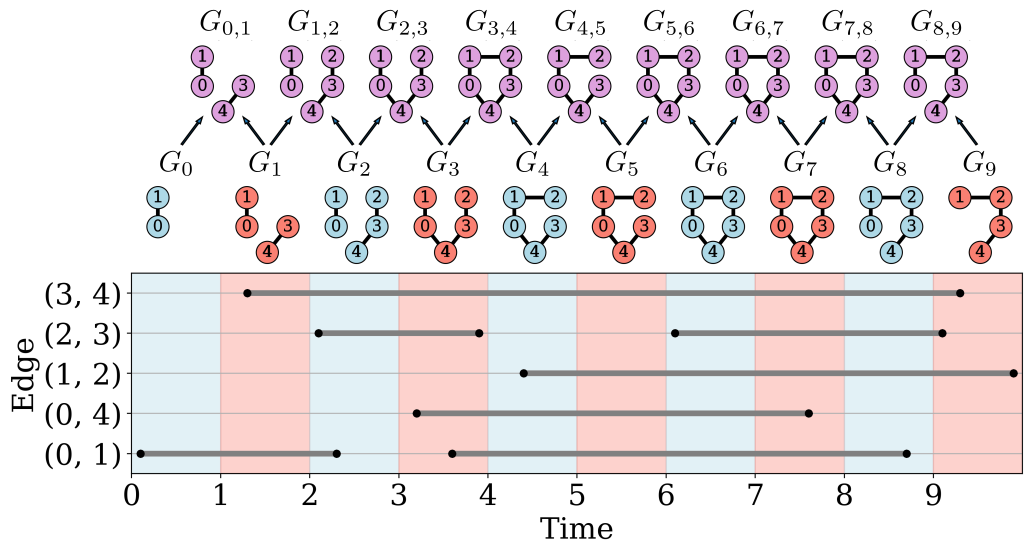}
     \end{minipage}
     \hfill
     \begin{minipage}{0.31\textwidth}
         \centering
         \includegraphics[width=\textwidth]{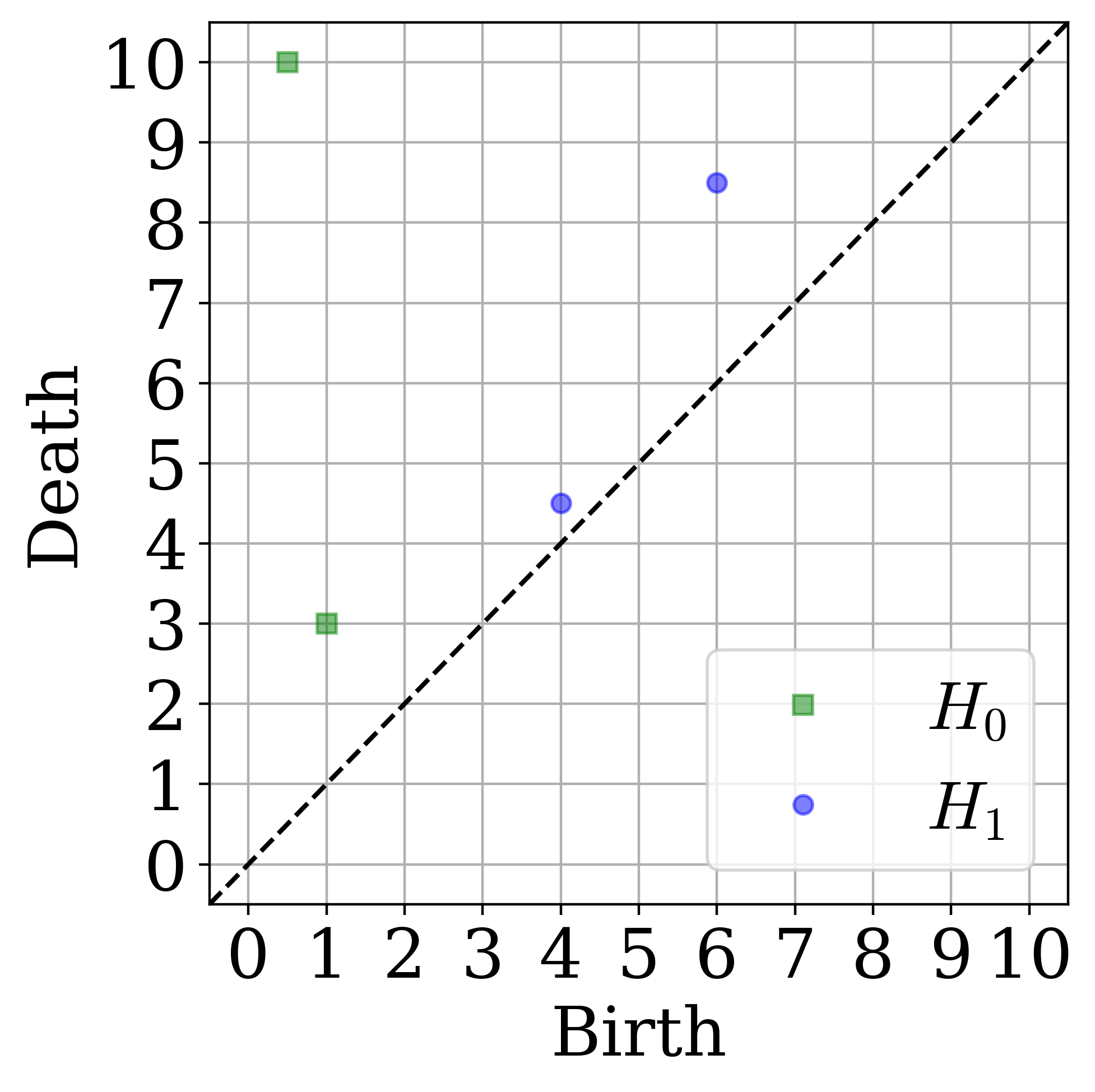} 
     \end{minipage}

     \caption{Example zigzag persistence applied to a simple temporal graph with temporal information stored for each edge as intervals. (Left) Edge intervals with sliding windows highlighted (alternating blue-red) with corresponding graphs and union graphs above. 
     (Right) Zigzag persistence diagram for both $H_0$ and $H_1$. The birth and death of a feature is encoded as the midpoint of the interval where the event happened. }
     \label{fig:example_zigzag_of_basic_temporal_graph}
\end{figure}

In the following simple example shown in Fig.~\ref{fig:example_zigzag_of_basic_temporal_graph}, we describe the method in more detail and show how to interpret the resulting zigzag persistence diagram.
In this example, we measure the changing structure of a simple 5-node cycle graph as edges are added and removed based on the temporal information. 
Fixing $r=1$, the simplicial complexes are exactly the same as the graphs $K_i = G_i$ as there are no cliques in this particular example.
The bottom left of Fig.~\ref{fig:example_zigzag_of_basic_temporal_graph}a shows the times associated to each edge, and the resulting graphs are shown above. 
In this notation, the subscripts correspond to the interval of time used to build the graphs. 
Then we have $w= 0.5$, and the centers of intervals are $t_0 = 0.5$, $t_1 = 1.5$, $\cdots$, $t_9 = 9.5$.
This results in intervals for graph $G_i$ of the form $(i,i+1)$ and intervals for the union graphs $G_{i,i+1}$ as $(i,i+2)$. 
At the end of the sliding windows, we consider the graph empty and set the death of any remaining homology features as the end time of the last window (i.e., $t=10$ for this example).

The resulting zigzag persistence diagram is shown in Fig.~\ref{fig:example_zigzag_of_basic_temporal_graph}b.
This persistence diagram shows the zero-dimensional and one-dimensional features as $H_0$ and $H_1$, respectively. There are two zero-dimensional features at persistence pairs $(1, 3)$ and $(0.5, 10)$. 
The later represents the connected component which appears at the first graph and lasts throughout the filtration. 
The other component is the piece consisting of vertices $3$ and $4$ which appears at union graph $G_{(0,2)}$, thus is associated to a birth at the midpoint of this interval occurring at 1.

The one-dimensional feature (the cycle represented in $H_1$) is present twice in the persistence diagram. 
This is due to it first appearing in $G_{(3,5)}$ and then disappearing at $G_{(4,5)}$ with corresponding persistence pair $(4, 4.5)$. 
The cycle then reappears at $G_{(5,7)}$ and disappears at $G_{(8,9)}$ resulting in persistence pair at $(6,8.5)$. 

This example demonstrates how zigzag persistence captures the changing structure of temporal graphs at multiple dimensions. 
It is possible to also capture higher-dimensional structures using higher-dimensional homology, although we do not investigate this direction in this work.
In particular, it is not clear what higher dimensional homology would represent in the context of data coming from 1-dimensional graph structures.

%% file: sections/sec-results.tex
\section{Results}
\label{sec:results}
To demonstrate the functionality of zigzag persistence for analyzing temporal graphs, we will use two examples. 
The first is an analysis of transportation data from Great Britain in Section~\ref{ssec:GB_results}. 
The second is a simulated dataset from the Lorenz system that exhibits intermittency, a dynamical system phenomenon where the dynamic state transition from periodic to chaotic in irregular intervals with results in Section~\ref{ssec:TOPN_results}. 
We study this signal using the temporal ordinal partition network framework as described in Section~\ref{ssec:temp_OPN}.
We compare our results for both examples to some standard networks tools to analyze temporal networks. Namely, we will compare two connectivity statistics and three centrality statistics. 

The two connectivity statistics analyze the Connected Components (CCs). The first CC statistic is the number of connected components $N_{cc}$, which provides a simple shape summary of the graph snapshots by understanding the number of disconnected subgraphs. The second statistic is the average size (number of nodes) of the connected components $\bar{S}_{cc}$. This statistic provides insight into how significant the components are for each graph snapshot.

The second statistic type is on centrality measures. The three centrality measures we use are the average and standardized degree centrality $\bar{C}_d$, betweenness centrality $\bar{C}_b$, and closeness centrality $\bar{C}_c$. 
The degree centrality measures the number of edges connected to a node, the betweenness centrality measures how often a node is used in all possible shortest paths, and the closeness centrality measures how close the node is to all other nodes through the shortest path.
For details on the implementation of each centrality measure, we direct the reader to~\cite{Landherr2010}.

\subsection{Great Britain Temporal Transportation Network} \label{ssec:GB_results}

\begin{figure}%
     \centering
     \begin{minipage}{0.31\textwidth}
         \centering
         \includegraphics[width=\textwidth]{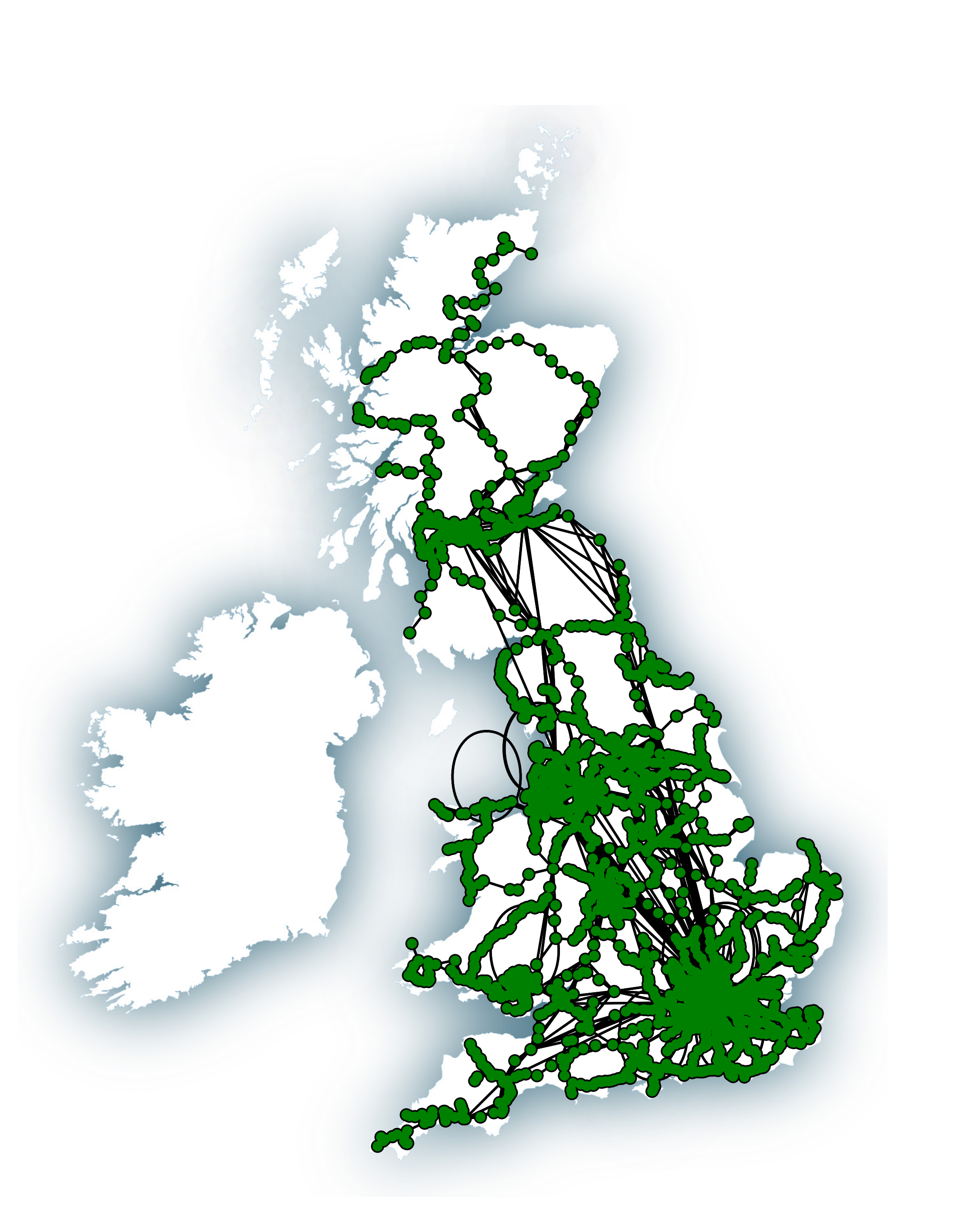} 
     \end{minipage}
     \hfill
     \begin{minipage}{0.31\textwidth}
         \centering
         \includegraphics[width=\textwidth]{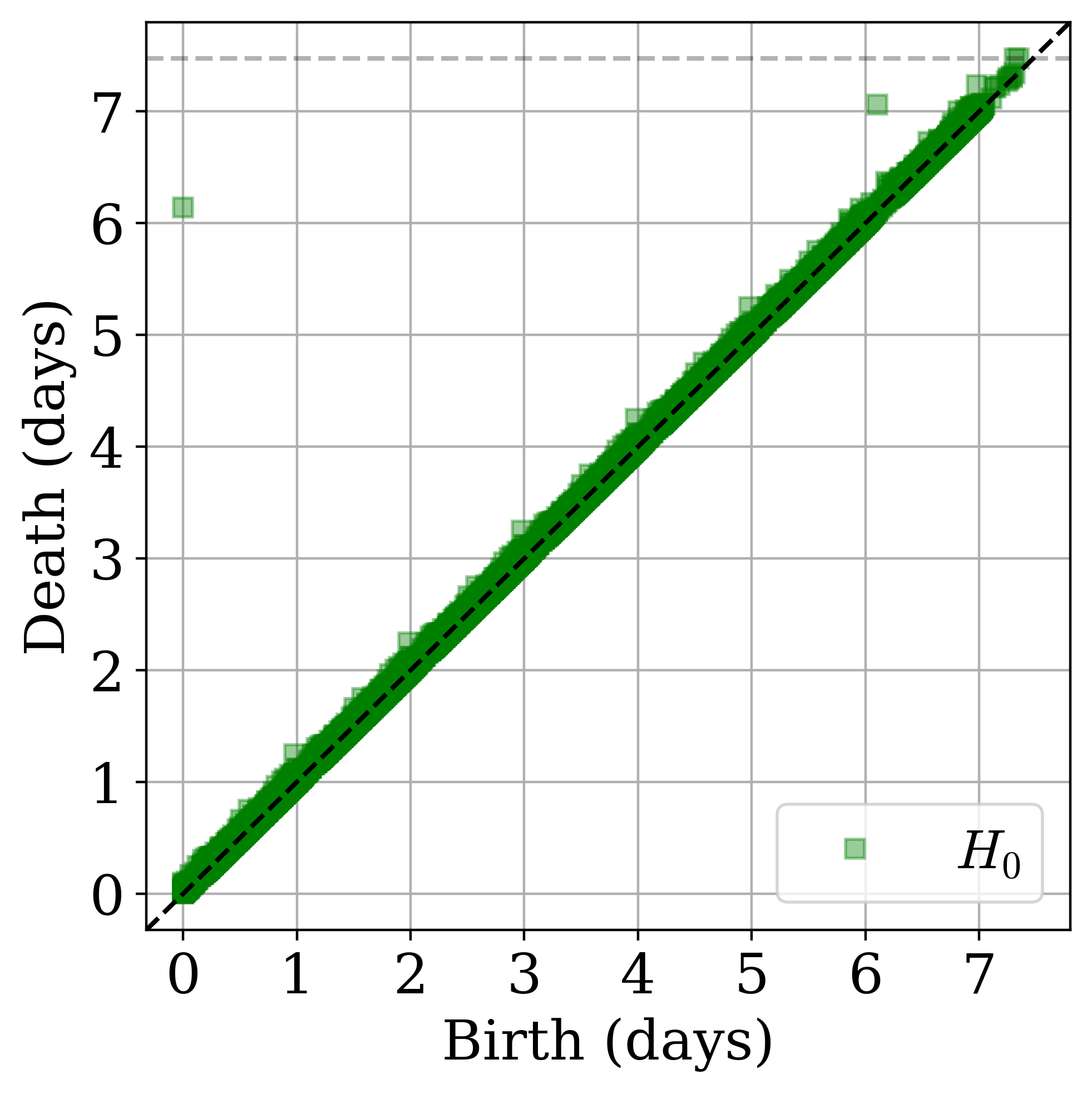} 
     \end{minipage}
     \hfill
     \begin{minipage}{0.31\textwidth}
         \centering
         \includegraphics[width=\textwidth]{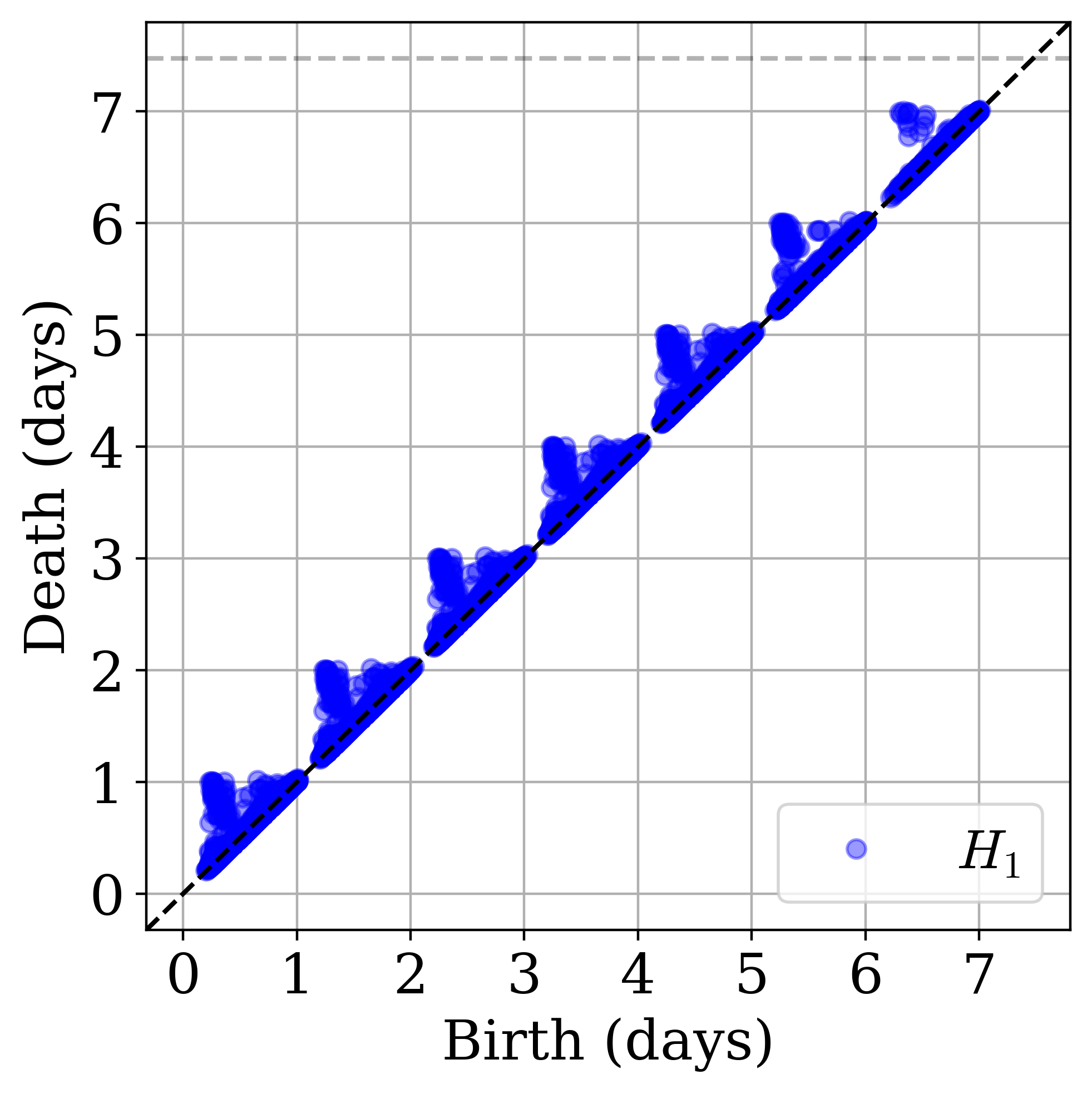}
     \end{minipage}

     \begin{minipage}{0.31\textwidth}
         \centering
         (a) Full Rail Travel Network.    
     \end{minipage}
     \hfill
     \begin{minipage}{0.31\textwidth}
         \centering
         (b) Zero-dimensional zigzag persistence.
     \end{minipage}
     \hfill
     \begin{minipage}{0.31\textwidth}
         \centering
         (c) One-dimensional zigzag persistence.
     \end{minipage}
     \caption{Zigzag persistence diagrams of the rail transportation network of Great Britain.}
     \label{fig:GB_rail_network_and_PDs}
\end{figure}
We use temporal networks created from the Great Britain (GB) temporal transportation dataset~\cite{Gallotti2015} for the air, rail, and coach transportation methods.
This data provides the destinations (nodes) and connections (edges) for public transportation in GB. 
Additionally, the departure and arrival times are provided to allow a temporal analysis. This temporal data was collected for one week, Monday through Sunday.
In this section, we use both the rail and coach data; similar calculations for air data are included in Appendix \ref{app:air_and_coach}.
The rail graph constructed without the inclusion of temporal information is shown at left in Fig.~\ref{fig:GB_rail_network_and_PDs} where the destinations are overlaid with a GB map outline. 
Figures for the similar air and coach graphs are included in Appendix \ref{app:air_and_coach}. 
In all three cases, we set the sliding windows to have width $w = 20$ minutes. 
Because the average wait time was 7 minutes and 7 seconds with a standard deviation of 7 minutes and 24 seconds from a collected sample~\cite{vanHagen2011}, this ensures that we retain connectivity. 
Additionally, we used an overlap of 50\% between adjacent windows.%

\subsubsection{Rail}
\begin{figure}
    \centering
    \includegraphics[width=0.99\textwidth]{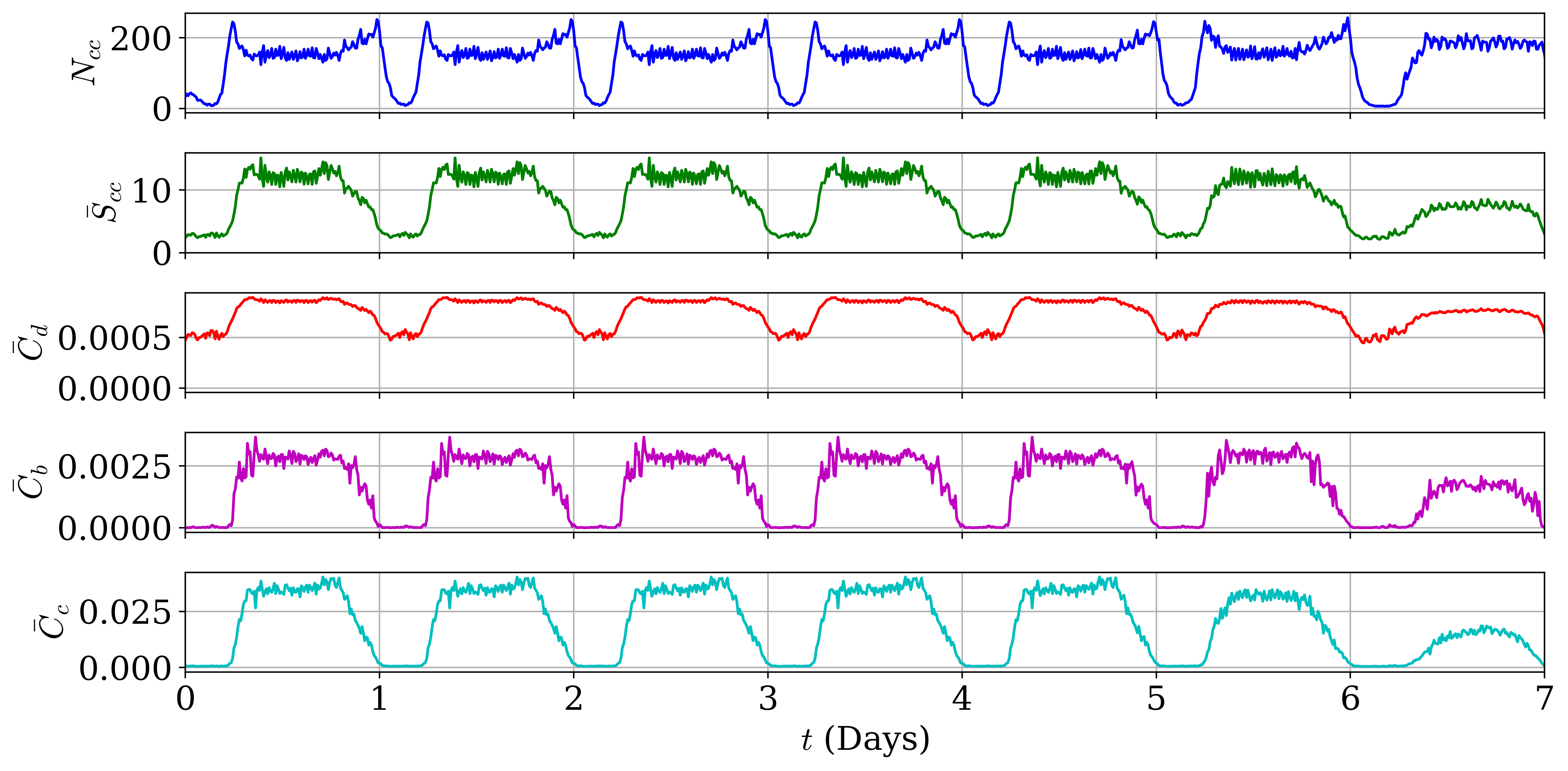}
    \caption{Connectivity and centrality analysis on temporal Great Britain rail network.}
    \label{fig:GB_rail_mean_centrality_components_analysis}
\end{figure}
We first focus on the rail data, for which we use the VR complex with $r=1$.
The 0- and 1-dimensional zigzag diagrams are shown in Fig.~\ref{fig:GB_rail_network_and_PDs}. 
The first noticeable feature is that the 1-dimensional diagram has a clear daily pattern of points, implying that there are many loops in the rail network that persist during the day, but become disconnected in the evenings when the trains are no longer running. 
However, there is still an overarching connectivity of the rail network seen during the week, as noted by the high persistence point at approximately $(0,6)$ in the 0-dimensional diagram. 
We note that this does not mean that the entire graph is connected for the entirety of the days Monday through Saturday, but instead that there is some connected area that remains running even overnight.
We suspect this would be tied to some of the more urban areas such as London, but further work would be needed to implement code to obtain generators of the zigzag diagram. 
This lower connectivity on Saturday and Sunday can also be seen in the additional sparsity in the day 6 and day 7 peaks in the 1-dimensional diagram. 

We compare these interpretations to the standard graph analysis tools \cite{Holme2015}. 
The standard centrality and connectivity statistics for the same data are shown in Fig.~\ref{fig:GB_rail_mean_centrality_components_analysis}.
We can clearly see that there is a daily periodicity in the data from these standard tools as was seen in the zigzag persistence. 
Specifically, all the connectivity and centrality measures increase during peak travel hours. 
What is lost, and which can be augmented with the zigzag viewpoint, is the ability to connect the clustering and centrality measures from one time step to the next. 
There is no way to determine, say, from the $N_{cc}$ graph that there is some portion of the graph that remains connected for 6 out of the 7 days. 
For this reason, we believe that the zigzag setting can be used along side the standard measures in order to strengthen analysis of given temporal graph information.

\begin{figure}%
     \centering
     \begin{minipage}{0.31\textwidth}
         \centering
         \includegraphics[width=\textwidth]{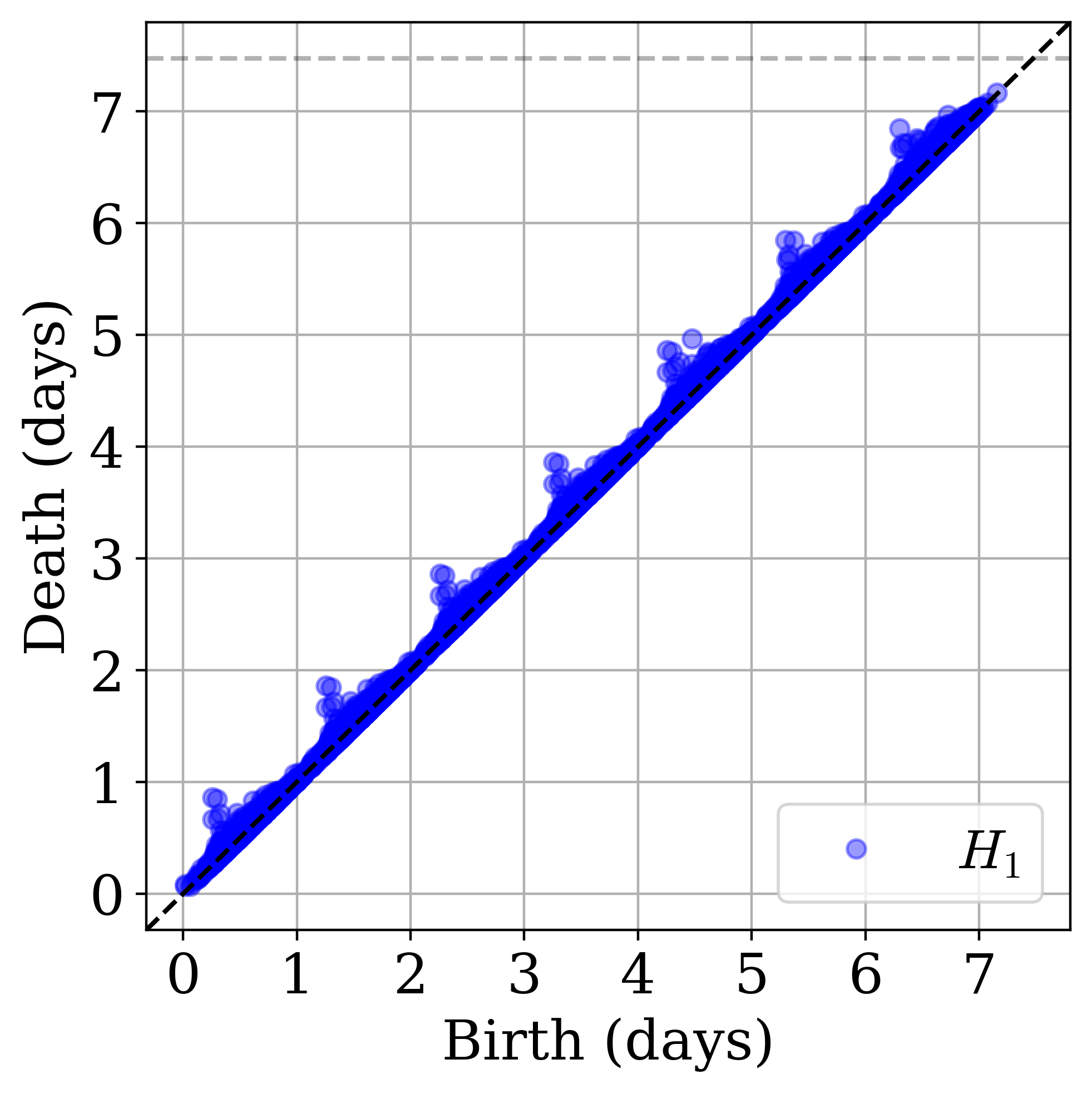}
     \end{minipage}
     \hfill
     \begin{minipage}{0.31\textwidth}
         \centering
         \includegraphics[width=\textwidth]{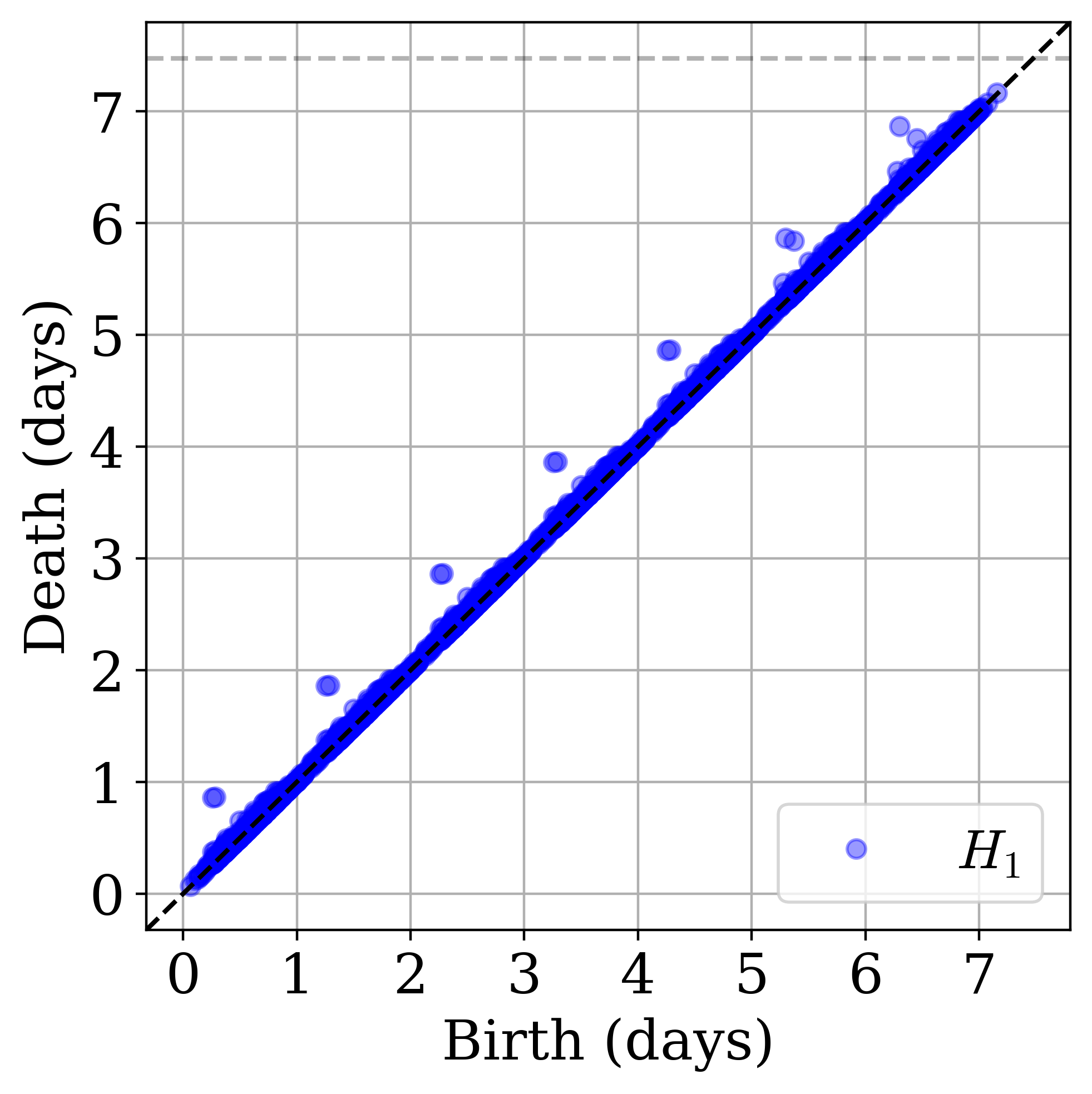}
     \end{minipage}
     \hfill
     \begin{minipage}{0.31\textwidth}
         \centering
         \includegraphics[width=\textwidth]{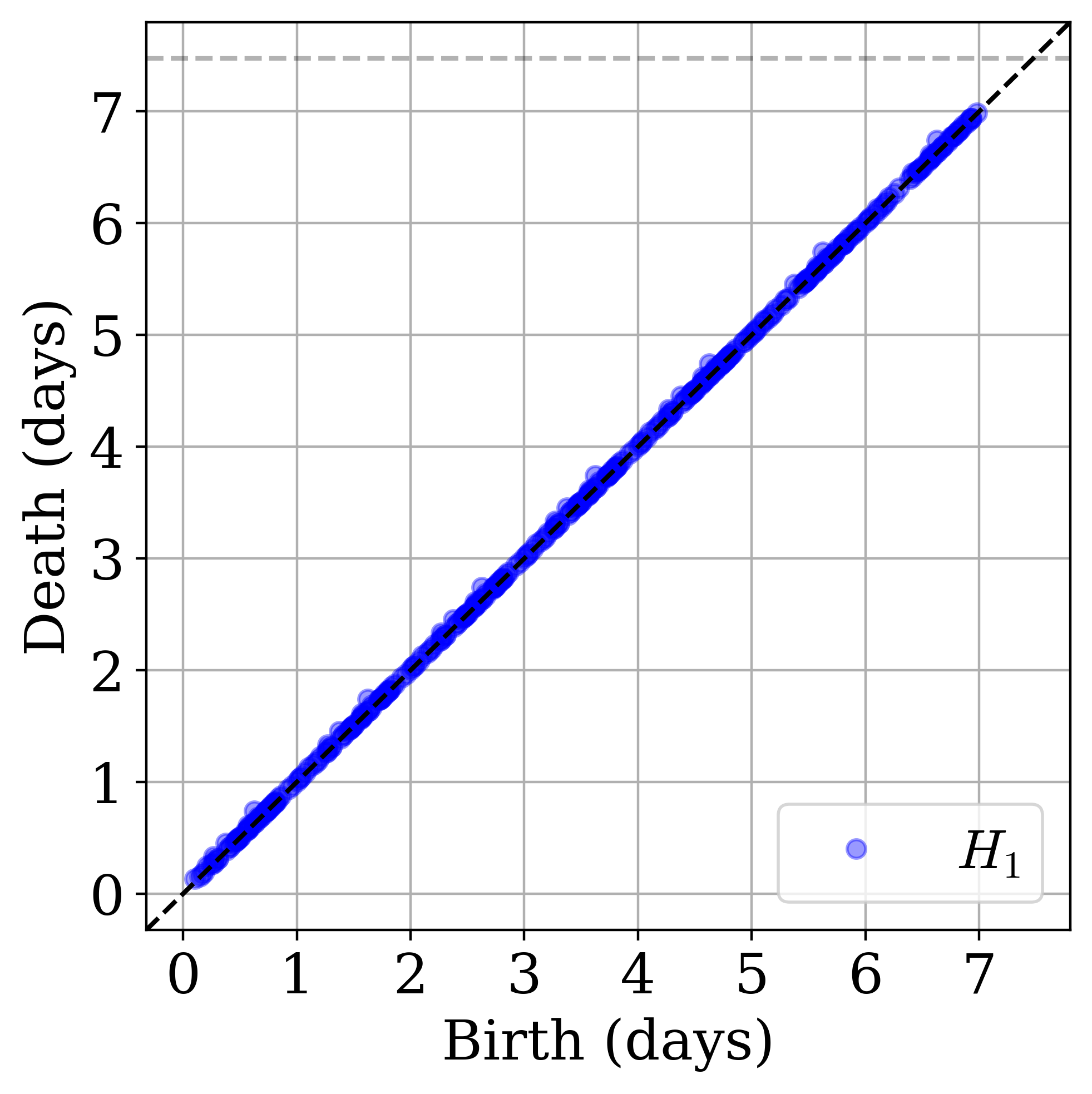}
     \end{minipage}

     \begin{minipage}{0.31\textwidth}
         \centering
         (a) One-dimensional zigzag persistence using original graph.
     \end{minipage}
     \hfill
     \begin{minipage}{0.31\textwidth}
         \centering
         (b) One-dimensional zigzag persistence using VR complex with $r=1$.
     \end{minipage}
     \hfill
     \begin{minipage}{0.31\textwidth}
         \centering
         (c) One-dimensional zigzag persistence using VR complex with $r=2$.
     \end{minipage}
     \caption{Zigzag persistence diagrams of the coach transportation network of Great Britain.}
     \label{fig:GB_coach_PD1s}
\end{figure}

\subsubsection{Coach}
\label{ssec:coach}
Now, we use the coach network data to illustrate using the VR complexes with different $r$ values. 
In Fig.~\ref{fig:GB_coach_PD1s}, 1-dimensional zigzag persistence diagrams are displayed. 
The leftmost diagram is computed using the graphs directly as input to the zigzag persistence without taking the VR complex.
The second and third were obtained using the graph and VR complexes with $r=1$ and $r=2$ respectively. 

Both diagrams (a) and (b) show periodic structure during the day which is to be expected as the number of buses running diminishes overnight. 
However, in diagram (b) we can clearly see a daily trend. 
While unfortunately difficult to see on the diagram, this is in fact points in the diagram overlaid, implying we have two loops of more than three edges that persist during the day. 
If the loops had length 3, they would be filled in by triangles when replacing the graph with the VR complex. 
This further implies that the noise seen in (a) is the existence of a great deal of triangles; not surprising in a highly connected local system like a bus network. 
But then, we can also compare persistence diagram Fig.~\ref{fig:GB_coach_PD1s}(b) to the analagous Fig.~\ref{fig:GB_rail_network_and_PDs}(c) from the rail network. 
The more locally connected nature of the bus system means that many of these small loops are not included when using the VR complex; however the rail system is more spread out, thus resulting in longer loops that are not filled in by a choice of $r=1$. 

Finally, we observe that in the persistence diagram of Fig.~\ref{fig:GB_coach_PD1s}(c), where there is very little of the periodic structure remaining. 
From this, we can assume that the many small loops seen in the $r=1$ diagram have been filled in by our choice of $r=2$, thus meaning that the loops are of length at most 6 (See the text surrounding Fig.6 of \cite{Myers2019} for a discussion of lifetime of loops by length in this setting).

\subsection{Temporal Ordinal Partition Network for Intermittency Detection} 
\label{ssec:TOPN_results}

In this section we apply the zigzag methods to study time series data encoded using complex networks; namely, the ordinal partition network.
Ordinal partition networks~\cite{McCullough2015} are a graph representation of time series data based on permutation transitions. 
As such, they encapsulate the state space structure of the underlying system. 
While we only use the ordinal partition network in this work, there are several other transitional complex networks from time-series data that a similar analysis could be done. These include $k$-nearest-neighbors~\cite{Khor2016}, epsilon-recurrence~\cite{Jacob2019}, and coarse-grained state-space networks~\cite{Small2009a,Small2013a}.
We begin by giving a brief introduction to the construction of the network from time series data, followed by the results of analysis using zigzag persistence.

\subsubsection{Temporal Ordinal Partition Network} 
\label{ssec:temp_OPN}
Given a time series $x = [x_0, x_1, x_2, \ldots, x_n]$ for a sequence of times $t = [t_0,\cdots, t_n]$, the ordinal partition network is formed by first generating a sequence of permutations  using a fixed permutation dimension $m$ and delay $\tau$. 
We generate a sequence of permutations by assigning each vector embedding 
\begin{equation}
v_i = [x_i, x_{i+\tau}, x_{i+2\tau}, \ldots, x_{i+(m-1)\tau}] = [v_i(0), v_i(1) \ldots, v_i(m-1)] 
\end{equation}
to one of the $m!$ possible permutations. 
We give an arbitrary ordering to the permutations for labeling purposes, and assign the permutation $\pi_j = [\pi_j(0), \ldots \pi_j(m-1)] \in \mathbb{Z}^{m}$ based on the ordinal pattern of $v_i$ such that 
$
v_i(\pi_j(0)) \leq v_i(\pi_j(1)) 
\leq 
\ldots \leq v_i(\pi_j(m-1))
$. 

Using the chronologically ordered sequence of permutations $\Pi$, we can form a graph $G(E,V)$ by setting the vertices $V$ as all permutations used and edges for transitions from $\pi_a$ to $\pi_b$ with $a, b \leq m!$ and $a \neq b$ (no self-loops). 
We will not add weight or directionality to the graph for this formation. 
However, we will track the index $i$ for the corresponding time $x_i$ when the edge is activated as the temporal data for the graph. 

\begin{figure}%
    \centering
    \includegraphics[width=0.82\textwidth]{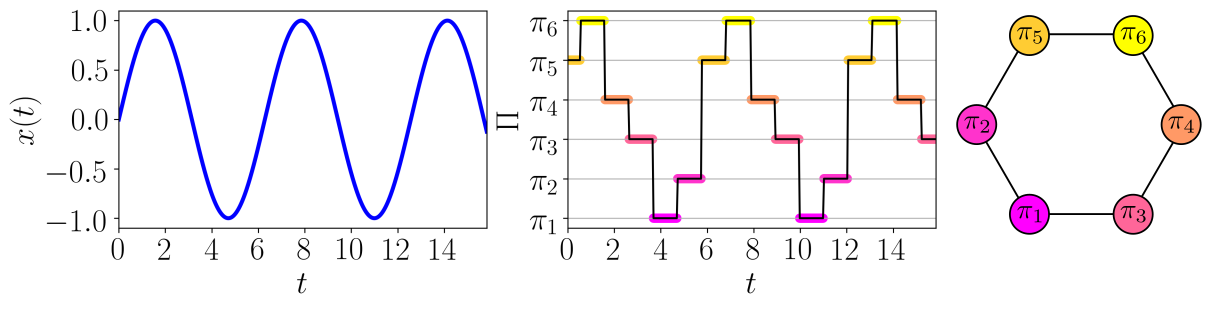}
    \caption{Example formation of an ordinal partition network for a sinusoidal signal $x(t) = \sin(t)$ with permutations of dimension $n=3$. The resulting permutation sequence $\Pi$ shows how these permutations transition, which is captured by the ordinal partition network on the right. 
    }
    \label{fig:simple_example_OPN_formation}
\end{figure}

In Fig.~\ref{fig:simple_example_OPN_formation} we demonstrate the ordinal partition network formation procedure for a simple example signal as $x(t) = \sin(t)$, where $t \in [0,15]$ sampled at a rate of $f_s = 25$ Hz. 
Using the method of multi-scale permutation entropy we selected $\tau = 52$ and set $n=3$ for demonstrative purposes. The corresponding permutations to delay embedding vector $v_i$ is shown as a sequence $\Pi$ in the middle subfigure of Fig.~\ref{fig:simple_example_OPN_formation}. This sequence captures the periodic nature of the signal which is then summarized as the ordinal partition network on the right with each permutation as a vertex and edges added for permutation transitions in $\Pi$.
For more details and examples of the ordinal partition network, we direct the reader to~\cite{McCullough2015, Myers2019}. 

\subsubsection{Ordinal Partition Network Results}
Using a sliding window technique, we can represent ordinal partition networks as temporal graphs. 
However, instead of each edge having a set of intervals associated with it as in the example in Section~\ref{ssec:example}, they have time instances where the edge is active  based on when a transition between unique permutations occur. 
For example, the transition from $\pi_i$ to $\pi_{i+1}$ occurring at time $t_j$ would be active for the moment in time $t_j$. 
If the sliding window contains an edge's activation instance, we add that edge to the sliding window graph.

We will show how this procedure can be used to detect chaotic and periodic windows in a signal exhibiting intermittency (i.e., the irregular transitions from periodic to chaotic dynamics). 
The signal used here is the $z$
solution to the simulated Lorenz system defined as
\begin{equation}
\frac{dx}{dt}   = \sigma (y-x), \: \frac{dy}{dt}   = x (\rho -z) - y, \: \frac{dz}{dt}   = xy - \beta z
 \label{eq:lorenz}
\end{equation}
with system parameters $\sigma = 10$, $\beta = 8/3$, and $\rho = 166.18$ for a response with type 1 intermittency~\cite{Pomeau1980}.
We simulated the system with a sampling rate of 100 Hz for 500 seconds with only the last 70 seconds used.
To construct the ordinal partition network, we choose $m=6$ and $\tau$ using the multi-scale permutation entropy method as suggested in~\cite{Myers2020c}. 
We set the sliding windows for generating graph snapshots to have a width of $5\tau$ and $80\%$ overlap between adjacent windows. 

\begin{figure}%
    \centering
    \includegraphics[width=0.99\textwidth]{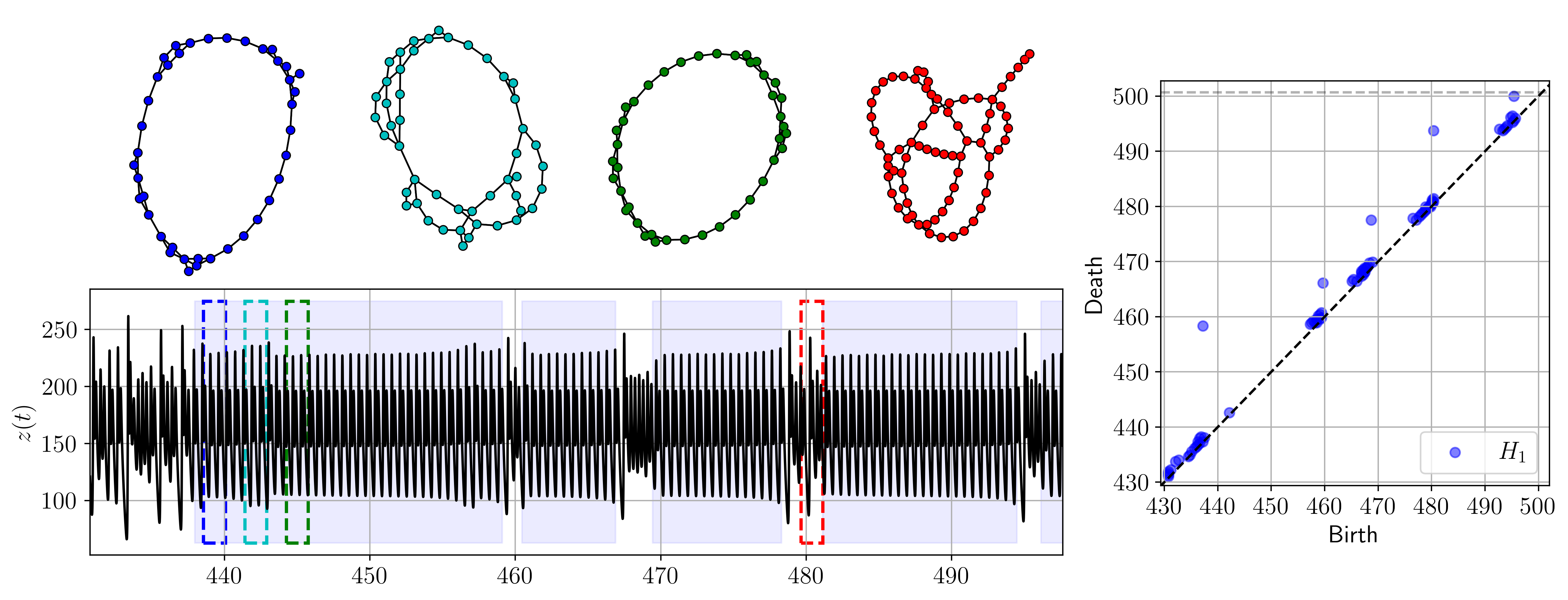}
    \caption{One the left, the $z$ solution of the intermittent Lorenz system described in Eq.~\eqref{eq:lorenz} is shown, along with four different graphs obtained from the corresponding ordinal partition networks in the windows of matching color. On the right, the one-dimensional zigzag persistence diagram.}
    \label{fig:intermittency_windows_and_PD1}
\end{figure}
The resulting signal $z(t)$ from simulating the Lorenz system in Eq.~\eqref{eq:lorenz} is shown in Fig.~\ref{fig:intermittency_windows_and_PD1}, with four examples of ordinal partition networks generated at several window locations. 
These sample graph snapshots show that the structure of the ordinal partition network significantly changes depending on the dynamic state of the window's time-series segment. 
At right in the figure, we see the 1-dimensional zigzag diagram for the data. 
Of particular note is that there are several high persistence points in the diagram. 
Recalling that the coordinates of these points are associated to time of appearance of loops rather than size of the loops themselves, we have marked the regions of the times series associated to these points in blue at left. 
From visual inspection, it appears that these regions correspond to periodic behavior in the time series. 
The remaining chaotic windows characteristically have many low-lifetime persistence pairs seen close to the diagonal in the persistence diagram. 
This is in line with the results in~\cite{Myers2019} that showed ordinal partition networks from chaotic signals tend to have persistence diagrams with many features in $H_1$ when compared to their periodic counterpart.
The fact that this labeling is done with only the user's choice of threshold for what is considered a high-persistence point makes this a potentially exciting avenue for future work to understand how it can be used in the case of labeling intermittency in time series. 

Additionally, note in the persistence diagram in Fig.~\ref{fig:intermittency_windows_and_PD1} the point near the diagonal, at around second 442, representing a loop in the network that quickly disappears, as seen also in the first three ordinal partition networks shown at left of Fig.~\ref{fig:intermittency_windows_and_PD1}. This tells us that low-lifetime loops may show up in the middle of periodic regions while the main periodic trend is preserved, confirming that we can still trust the coordinates of high persistence points as the bounds of periodic regions.
Thus, in general, we can use the persistence diagram to identify periodic regions by looking for intervals where there are persistent loops over relatively long periods and very few (or no) other shorter-living loops, which is easily identified from the persistent points coordinates.

\begin{figure}%
    \centering
    \includegraphics[width=0.99\textwidth]{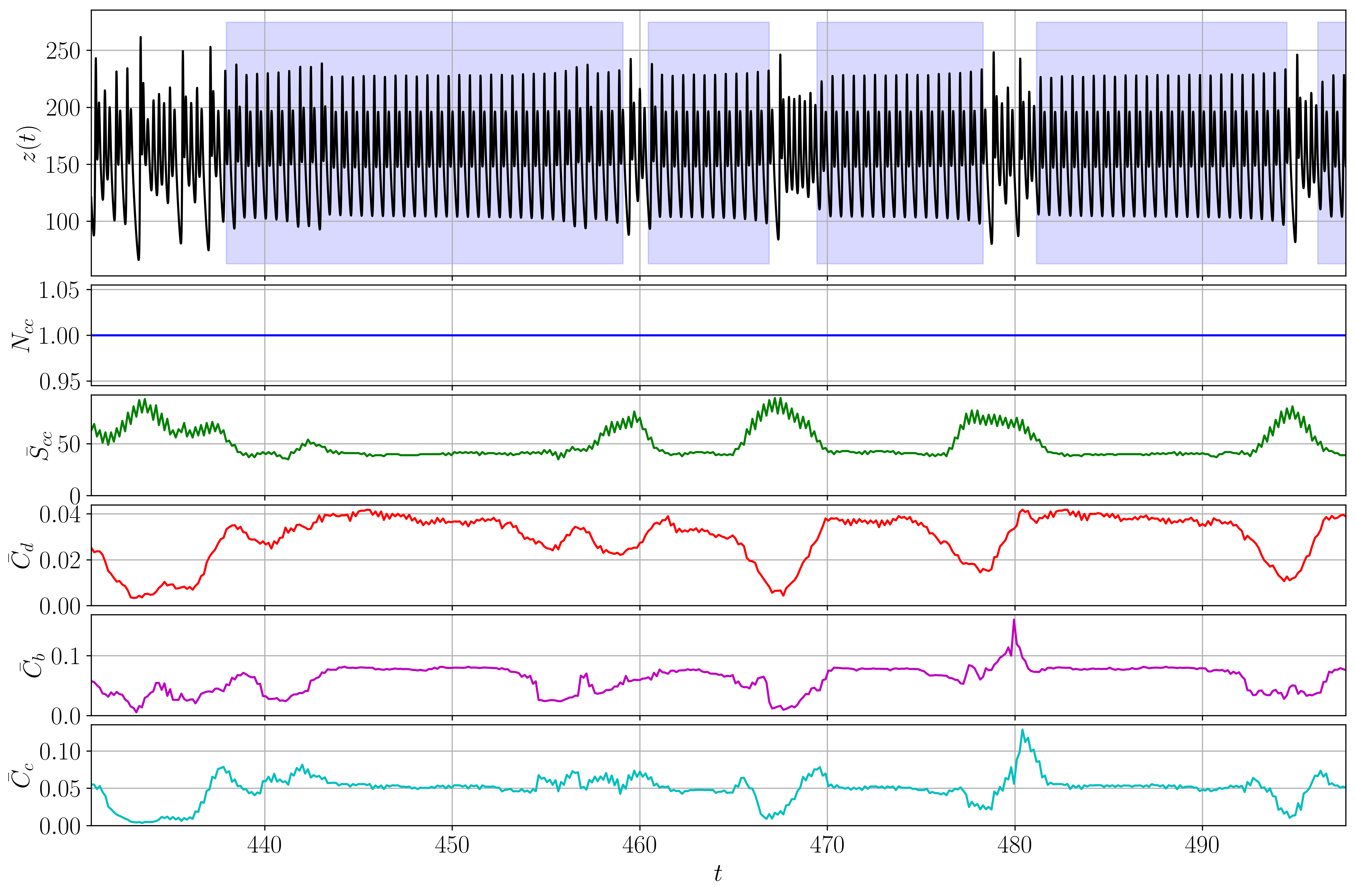}
    \caption{Connectivity and centrality analysis on temporal ordinal partition network with periodic regions as labeled with the zigzag persistence points highlighted in blue.}
    \label{fig:intermittency_mean_centrality_components_analysis}
\end{figure}

To compare with the standard tools from temporal network analysis, we show the connectivity and centrality measures of the graph snapshots in Fig.~\ref{fig:intermittency_mean_centrality_components_analysis}. 
The number of components $N_{cc}$ is constant due to the nature of the ordinal partition network, where the sequence of permutation transitions creates a chain of connected edges. 
As such, there is no structural information in the number of components. 
However, the size of the components does increase during the chaotic windows. This increase is due to, in general, more unique permutations and thus nodes used in a chaotic signal compared to periodic.
Of the centrality statistics, only the average closeness centrality shows an apparent increase during chaotic regions. 
The increase in centrality is most likely due to the chaotic regions causing a more highly connected graph as demonstrated in the chaotic window and corresponding network of Fig.~\ref{fig:intermittency_mean_centrality_components_analysis}.
While these statistics do provide some insight into the changing dynamics, they do not show how the higher-dimensional structure of the graph evolves through the sliding windows and graph snapshots, in contrast to zigzag persistence.

%% file: sections/sec-conclusion.tex
\section{Conclusion}
\label{sec:conclusion}
In this work we investigated one method for applying zigzag persistence to temporal graphs and discussed interpretation in several settings. 
We treated these graphs as inputs to create a metric space, and then studied the zigzag constructed by the changing graph over time for a fixed connectivity parameter. 
Zigzag persistence provides a unique perspective when studying the evolving structure of a temporal graph by tracking the standard lower-dimensional features (e.g., connected components), but also higher-dimensional features (e.g., loops and voids) through a sequence of simplicial complexes.
This allows for an understanding of the evolving topology of a temporal graph which can be used in addition to more standard techniques for their analysis.
We showed interpretation for these tools on two examples: the Great Britain transportation network and temporal ordinal partition networks. 
Our results showed that the informative zero and one-dimensional zigzag persistence provided insights into the structure of the temporal graph that were not easily gleaned from standard centrality and connectivity statistics. 

It should be noted that the work here is essentially functioning as a more computationally tractable version of what we would like to do, namely two-parameter persistence. 
Because multiparameter persistence presents mathematical barriers to simplified representations in the spirit of persistence diagrams \cite{Carlsson2009a}, many other workarounds have sprung up. 
These include vineyards \cite{CohenSteiner2006}, where we study a one parameter family of persistence diagrams; or CROCKER plots \cite{Ulmer2019,Guezel2022}, where we can study a similarly evolving Betti curve. 
\revision{ 
There has been previous work investigating the use of zigzag persistence in graph based applications in the broader theoretical framework of \textit{formigrams} \cite{Kim2022,Kim2019} with additional results on stability \cite{Kim2020}.
These include applications to dynamic metric spaces \cite{Kim2020a, Kim2021} and brain networks \cite{Chowdhury2018c}, although this setting is restricted to 0-dimensional persistence due to its tight connection to time-varying clustering.}
\revisionRemove{Until recently, there has been limited involvement of zigzag persistence in these types of applications since e}Even though in theory the running time \revision{of zigzag persistence} should be similar to standard persistence \cite{Milosavljevic2011}, in practice it has not seen the flurry of optimizations available in the regular case \cite{Bauer2021,Otter2017}. 
However, recent work \cite{Dey2021a,Dey2022,Dey2021b} promises substantial improvements in the potentially available code, which should further make the tools discussed in this paper more accessible to a wide array of data sets.

We believe zigzag persistence could also be leveraged to study other temporal graphs including flock behavior models (e.g., Viscsek model) and the emergence of coordinated motion, power grid dynamics with the topological characteristics of a cascade failures, and supplier-manufacture networks through the effects of trade failures on production and consumption.
Future work could involve an analysis on deciding an optimal window size and overlap, a method to incorporate edge weight and directionality, and temporal information on both the nodes and edges. 
It would also be worth investigating higher-dimensional features (e.g., voids through $H_2$), although more likely because of the 1-dimensional structure of the input data, it is less clear what sorts of behavior might arise in the form of interesting behavior in the zigzag diagrams.

Additionally, the method presented here is by no means the only way to incorporate the input temporal graph in a form that could generate a zigzag complex. 
We direct the interested reader to \cite{Aktas2019} for a large collection of possible 1-parameter filtrations that could be generated from a fixed graph. 
Utilizing any of these for some fixed filtration parameter evolving over the temporal graph information could yield a zigzag diagram whose structure may be useful for additional interpretation in other applications.

%% file: sections/sec-acknowledgements.tex
This material is based upon work supported by the Air Force Office of Scientific Research under award number FA9550-22-1-0007.

%% file: sections/sec-appendix.tex
\section{Great Britain Transportation Networks Results: Air and Coach Travel} 
\label{app:air_and_coach}

For completeness, we include figures and persistence diagrams for the Great Britain transportation data not included in Sec.~\ref{ssec:GB_results}. 

\subsection{Coach Travel Network Analysis}
\begin{figure}
     \centering

     \begin{minipage}{0.3\textwidth}
         \centering
         \includegraphics[width=\textwidth]{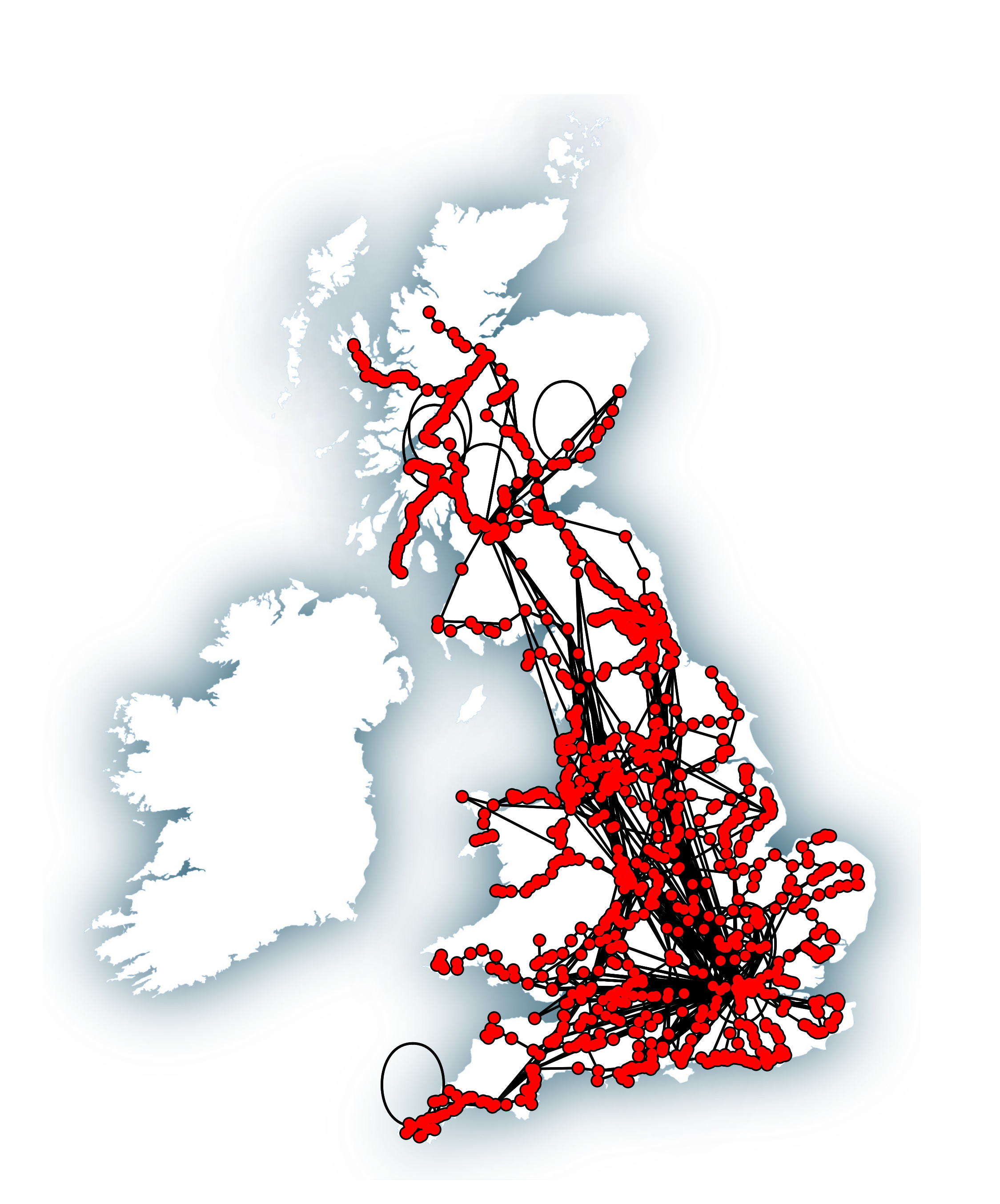} \\
         (a) Full coach Travel Network.
     \end{minipage}
     \hfill
     \begin{minipage}{0.3\textwidth}
         \centering
         \includegraphics[width=\textwidth]{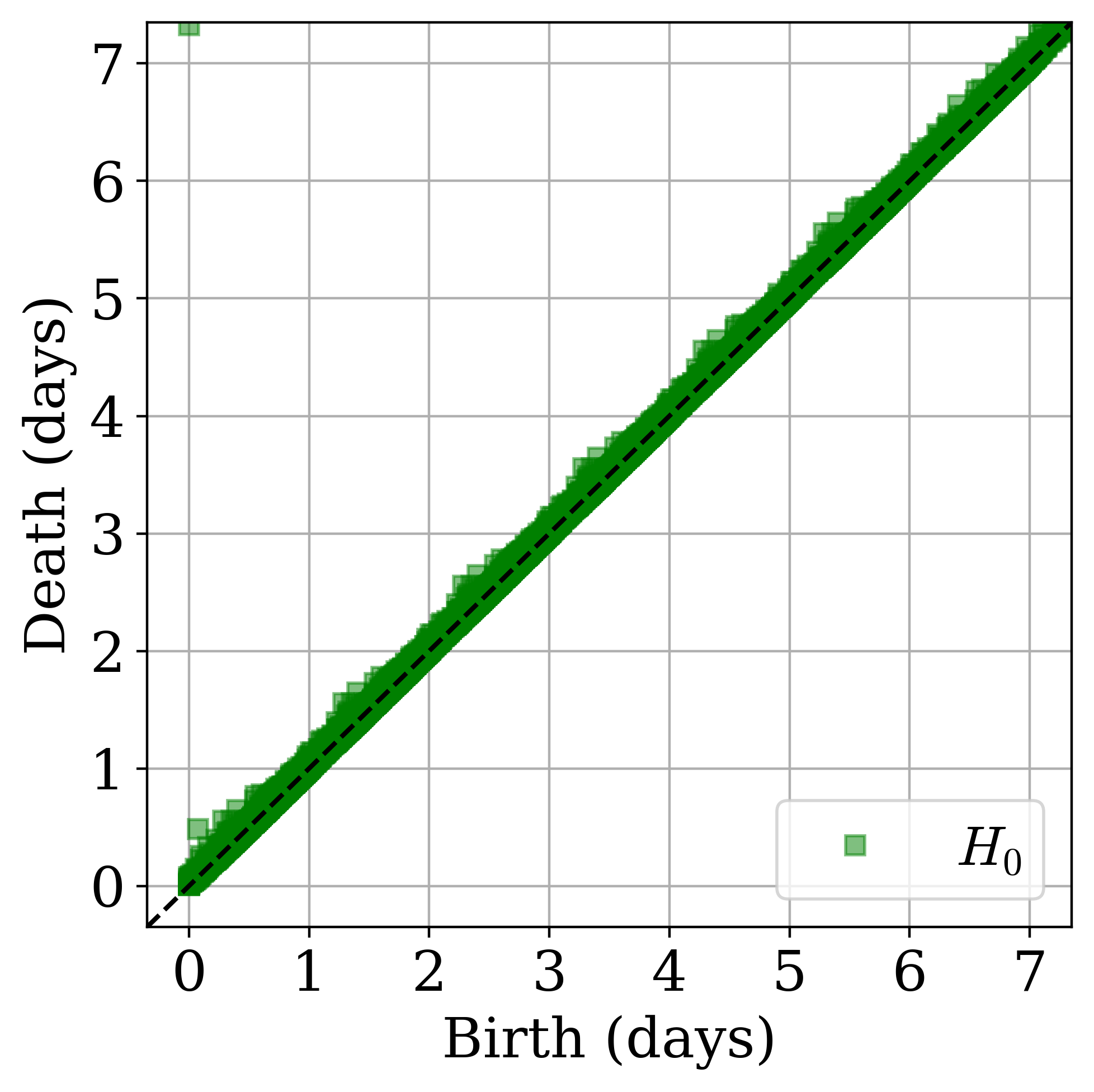} \\
         (b) Zero-dimensional zigzag persistence.
     \end{minipage}
     \hfill
     \begin{minipage}{0.3\textwidth}
         \centering
         \includegraphics[width=\textwidth]{Figures/GB_coach_PD1_Rips1} \\
         (c) One-dimensional zigzag persistence.
     \end{minipage}

     \qquad 
     
    \includegraphics[width=0.99\textwidth]{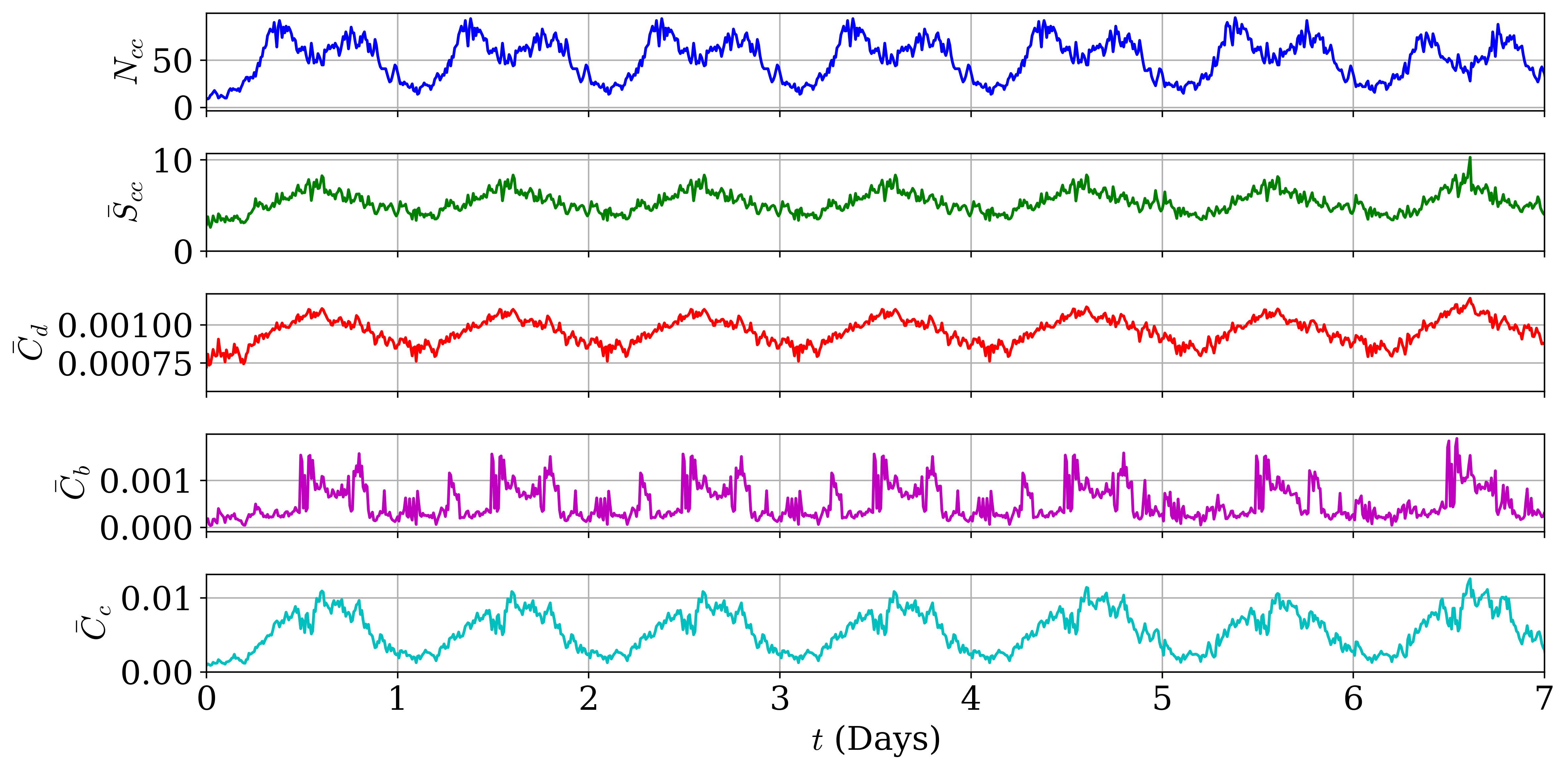}
     (d) Connectivity and centrality analysis on temporal Great Britain coach network.
     \caption{Results for the coach transportation network of Great Britain.}
     \label{fig:GB_coach_all}
\end{figure}

Figure~\ref{fig:GB_coach_all} shows the full network for the coach data, and the 0- and 1-dimensional zigzag diagrams in the top row. 
The standard statistics for temporal graphs are shown in the bottom row. 
See Sec.~\ref{ssec:coach} for a full discussion of the coach data. 
\subsection{Air Travel Network Analysis}
\begin{figure}
     \centering
     \begin{minipage}{0.3\textwidth}
         \centering
         \includegraphics[width=\textwidth]{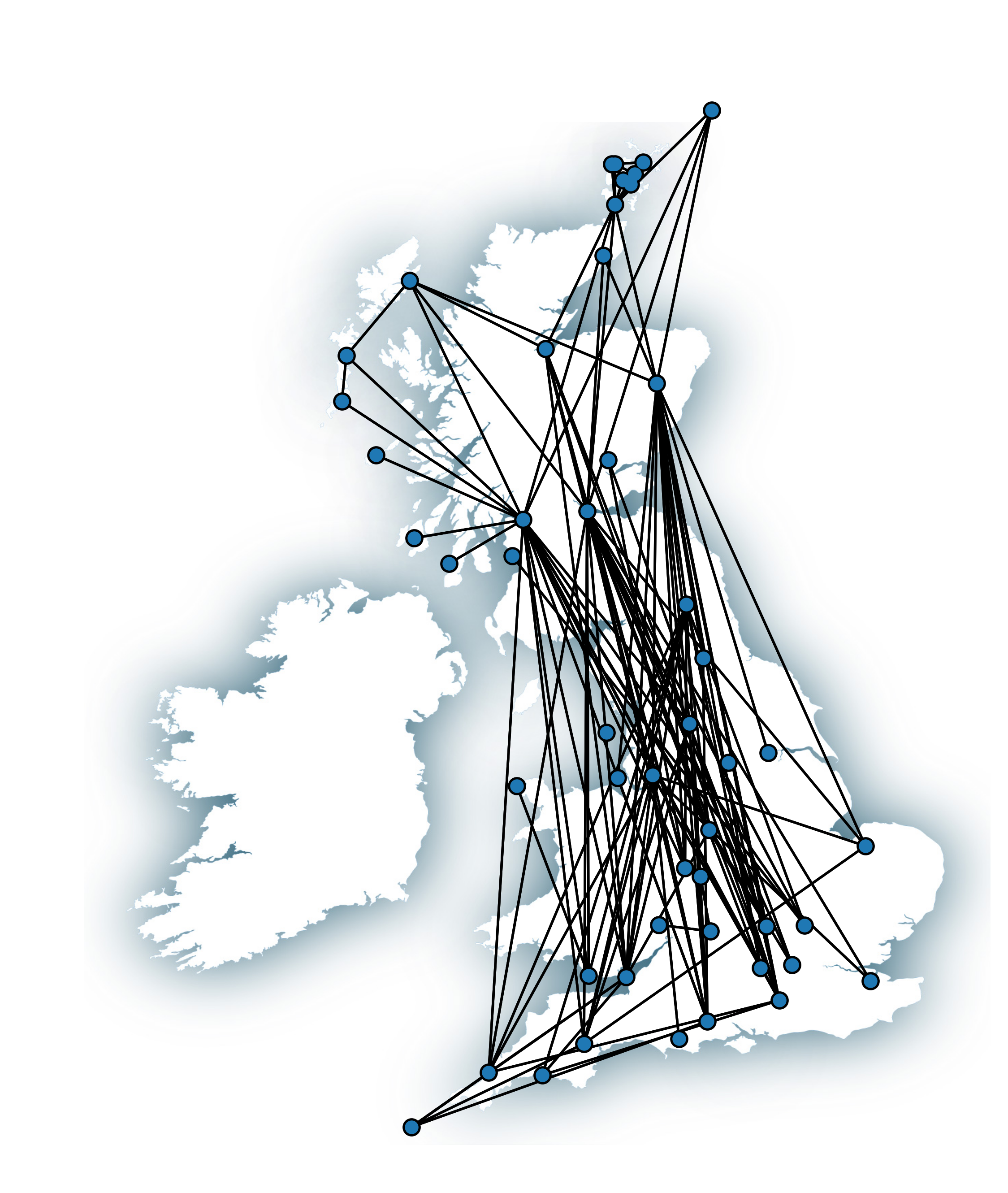} \\
         (a) Full air Travel Network.
     \end{minipage}
     \hfill
     \begin{minipage}{0.3\textwidth}
         \centering
         \includegraphics[width=\textwidth]{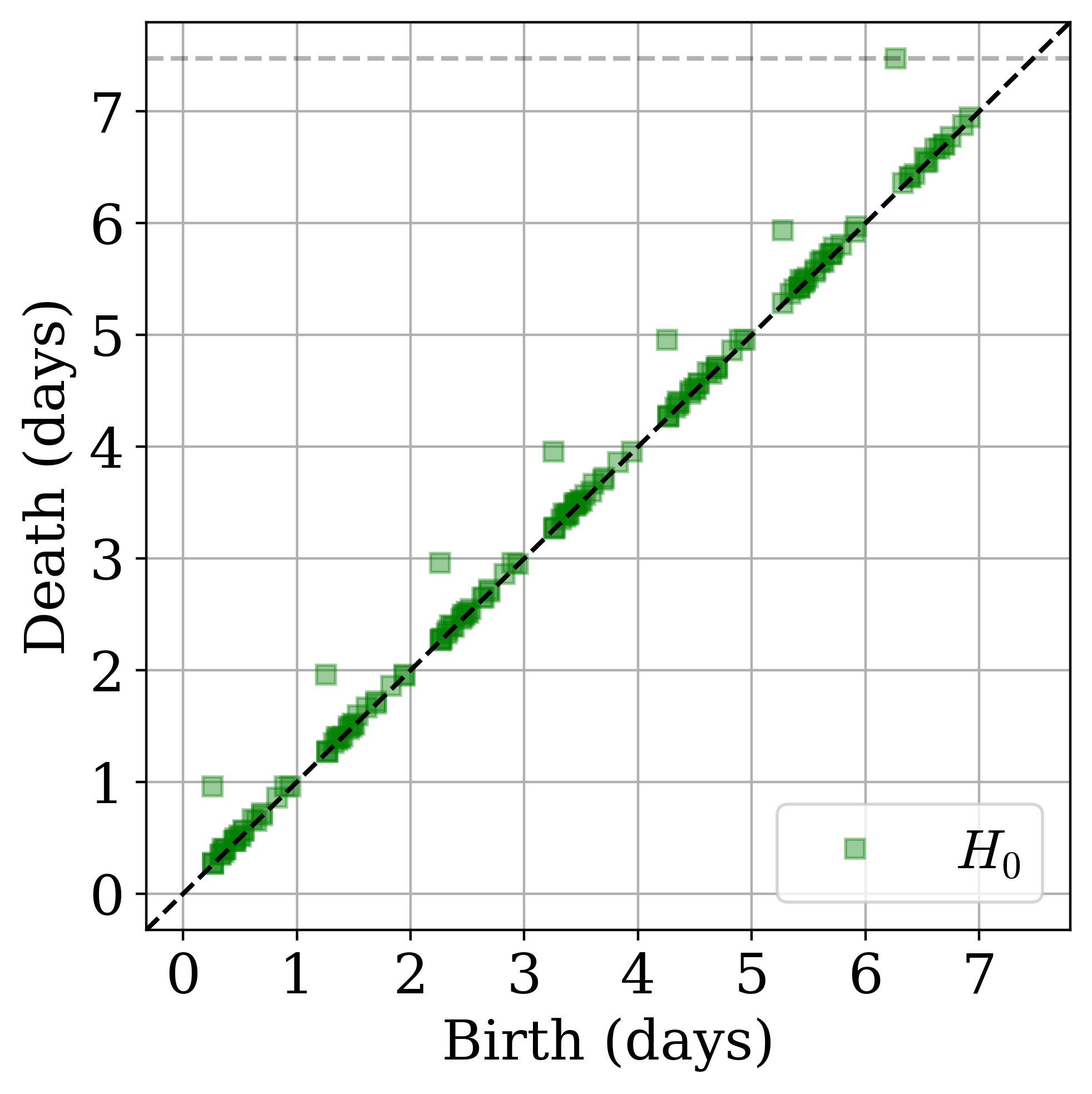} \\
         (b) Zero-dimensional zigzag persistence.
     \end{minipage}
     \hfill
     \begin{minipage}{0.3\textwidth}
         \centering
         \includegraphics[width=\textwidth]{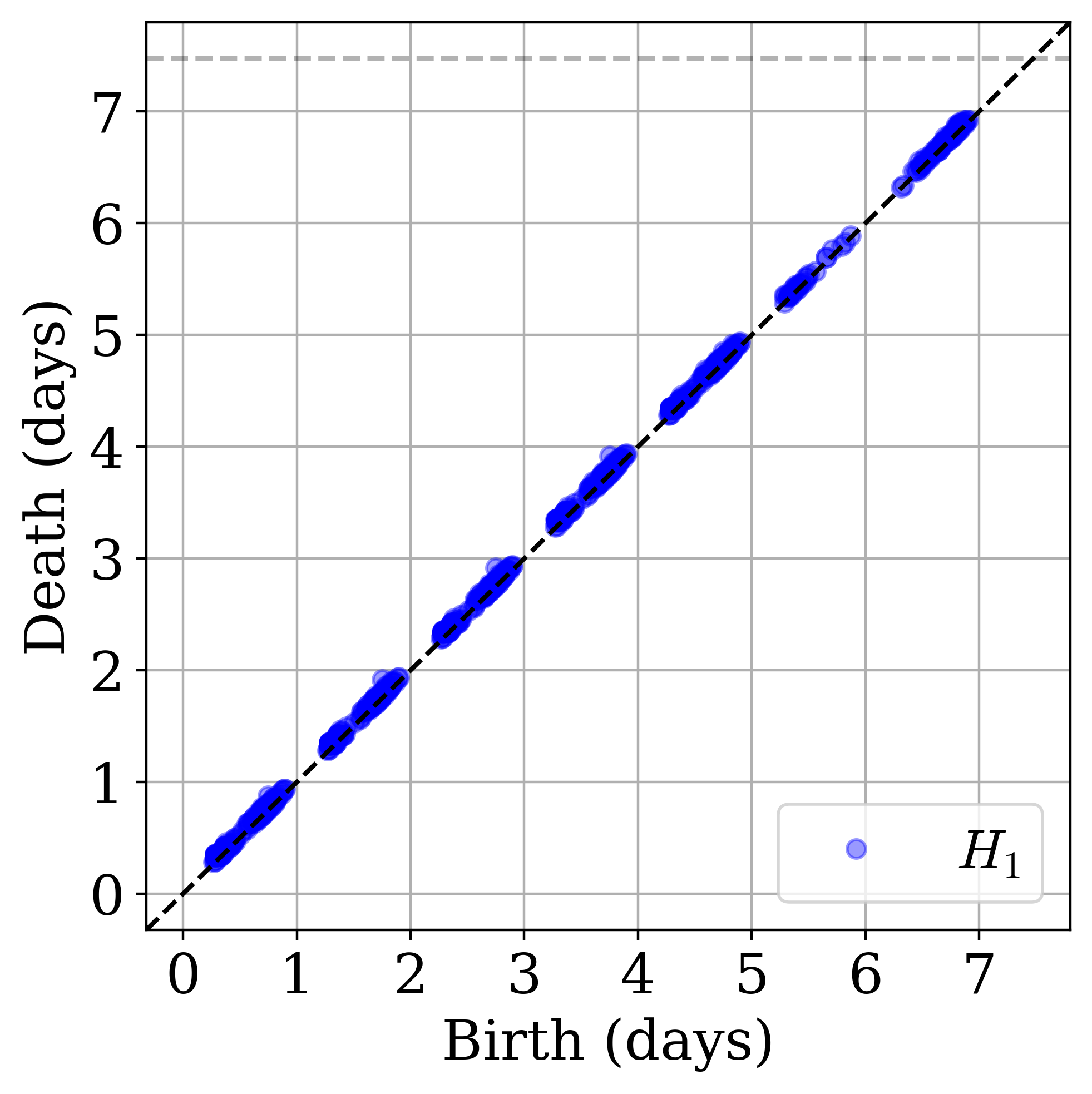} \\
         (c) One-dimensional zigzag persistence.
     \end{minipage}

     \qquad 
     
    \includegraphics[width=0.99\textwidth]{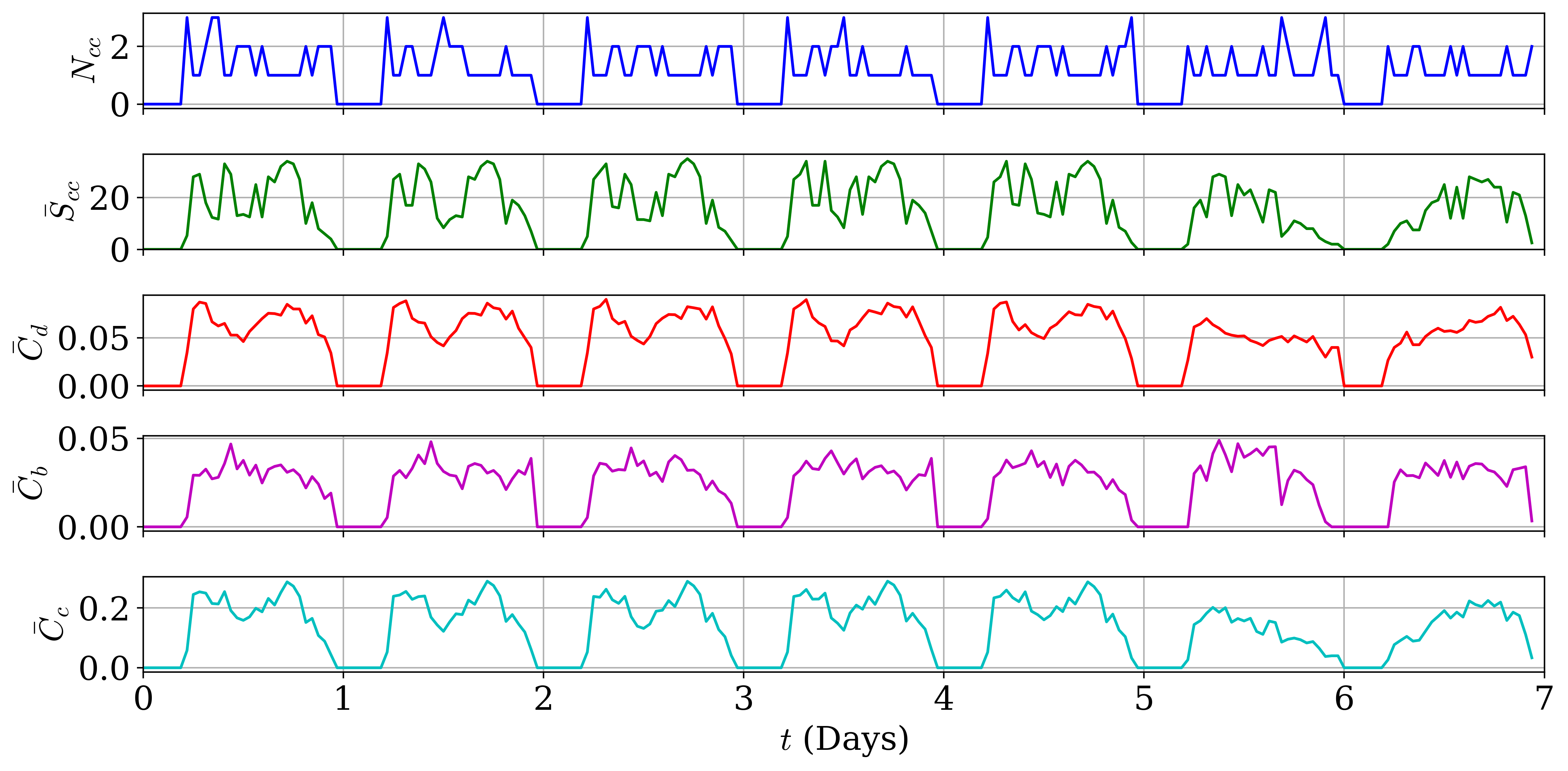}
    (d)Connectivity and centrality analysis on temporal Great Britain air network.
     \caption{Results for the air transportation network of Great Britain.}
     \label{fig:GB_air_all}
\end{figure}

Figures~\ref{fig:GB_air_all} shows the zigzag diagrams and standard measures for the air travel data. 
Like rail and coach, the air travel network clearly has regular daily structure as seen in both diagrams. 
Interestingly, here we have a daily persistence point in 0-dimensions with splits in the middle, meaning that the air network does not retain any connected regions overnight. 
This, again, is not surprising as there is more tendency for airports to have absolutely no flights overnight as opposed to late-night bus systems. 
The lack of high persistence points in $H_1$ suggests that there are not even many small loops in the network at any given time, which might be caused by having considerably fewer edges in this network than the others.

%% file: arXiv_PH_of_DN_article.bbl
\begin{thebibliography}{10}

\bibitem{Porter2020}
Mason~A. Porter.
\newblock Nonlinearity + networks: A 2020 vision.
\newblock In {\em Emerging Frontiers in Nonlinear Science}, pages 131--159.
  Springer International Publishing, 2020.

\bibitem{Holme2013}
Petter Holme and Jari Saramäki, editors.
\newblock {\em Temporal Networks}.
\newblock Springer Berlin Heidelberg, 2013.

\bibitem{Holme2015}
Petter Holme.
\newblock Modern temporal network theory: a colloquium.
\newblock {\em The European Physical Journal B}, 88(9), sep 2015.

\bibitem{Skyrms2000}
B.~Skyrms and R.~Pemantle.
\newblock A dynamic model of social network formation.
\newblock {\em Proceedings of the National Academy of Sciences},
  97(16):9340--9346, aug 2000.

\bibitem{Husein2019}
Ismail Husein, Herman Mawengkang, Saib Suwilo, and Mardiningsih.
\newblock Modeling the transmission of infectious disease in a dynamic network.
\newblock {\em Journal of Physics: Conference Series}, 1255(1):012052, aug
  2019.

\bibitem{Xu2019b}
Mengkai Xu, Srinivasan Radhakrishnan, Sagar Kamarthi, and Xiaoning Jin.
\newblock Resiliency of mutualistic supplier-manufacturer networks.
\newblock {\em Scientific Reports}, 9(1), sep 2019.

\bibitem{Schaefer2018}
Benjamin Schäfer, Dirk Witthaut, Marc Timme, and Vito Latora.
\newblock Dynamically induced cascading failures in power grids.
\newblock {\em Nature Communications}, 9(1), may 2018.

\bibitem{DavidBoyce2012}
Bin~Ran David~Boyce.
\newblock {\em Modeling Dynamic Transportation Networks}.
\newblock Springer Berlin Heidelberg, 2012.

\bibitem{Enright2018}
Jessica Enright and Rowland~Raymond Kao.
\newblock Epidemics on dynamic networks.
\newblock {\em Epidemics}, 24:88--97, sep 2018.

\bibitem{Nuss2016}
Philip Nuss, T.E. Graedel, Elisa Alonso, and Adam Carroll.
\newblock Mapping supply chain risk by network analysis of product platforms.
\newblock {\em Sustainable Materials and Technologies}, 10:14--22, dec 2016.

\bibitem{Soltan2014}
Saleh Soltan, Dorian Mazauric, and Gil Zussman.
\newblock Cascading failures in power grids.
\newblock In {\em Proceedings of the 5th international conference on Future
  energy systems}. {ACM}, jun 2014.

\bibitem{Bast2015}
Hannah Bast, Daniel Delling, Andrew Goldberg, Matthias Müller-Hannemann,
  Thomas Pajor, Peter Sanders, Dorothea Wagner, and Renato~F. Werneck.
\newblock Route planning in transportation networks, 2015.

\bibitem{Sugishita2020}
Kashin Sugishita and Yasuo Asakura.
\newblock Vulnerability studies in the fields of transportation and complex
  networks: a citation network analysis.
\newblock {\em Public Transport}, 13(1):1--34, sep 2020.

\bibitem{Hackl2019}
Jürgen Hackl and Bryan~T. Adey.
\newblock Estimation of traffic flow changes using networks in networks
  approaches.
\newblock {\em Applied Network Science}, 4(1), may 2019.

\bibitem{Chen2016a}
Xiaoying Chen, Chong Zhang, Bin Ge, and Weidong Xiao.
\newblock Temporal query processing in social network.
\newblock {\em Journal of Intelligent Information Systems}, 49(2):147--166, dec
  2016.

\bibitem{Huang2015}
Silu Huang, Ada Wai-Chee Fu, and Ruifeng Liu.
\newblock Minimum spanning trees in temporal graphs.
\newblock In {\em Proceedings of the 2015 {ACM} {SIGMOD} International
  Conference on Management of Data}. {ACM}, may 2015.

\bibitem{Wang2019}
Yishu Wang, Ye~Yuan, Yuliang Ma, and Guoren Wang.
\newblock Time-dependent graphs: Definitions, applications, and algorithms.
\newblock {\em Data Science and Engineering}, 4(4):352--366, sep 2019.

\bibitem{Borgatti2005}
Stephen~P. Borgatti.
\newblock Centrality and network flow.
\newblock {\em Social Networks}, 27(1):55--71, jan 2005.

\bibitem{Crawford2018}
Joseph Crawford and Tijana Milenkovi{\'{c}}.
\newblock {ClueNet}: Clustering a temporal network based on topological
  similarity rather than denseness.
\newblock {\em {PLOS} {ONE}}, 13(5):e0195993, may 2018.

\bibitem{You2021}
Jingyi You, Chenlong Hu, Hidetaka Kamigaito, Kotaro Funakoshi, and Manabu
  Okumura.
\newblock Robust dynamic clustering for temporal networks.
\newblock In {\em Proceedings of the 30th {ACM} International Conference on
  Information {\&} Knowledge Management}. {ACM}, oct 2021.

\bibitem{Moriano2019}
Pablo Moriano, Jorge Finke, and Yong-Yeol Ahn.
\newblock Community-based event detection in temporal networks.
\newblock {\em Scientific Reports}, 9(1), mar 2019.

\bibitem{Kempe2002}
David Kempe, Jon Kleinberg, and Amit Kumar.
\newblock Connectivity and inference problems for temporal networks.
\newblock {\em Journal of Computer and System Sciences}, 64(4):820--842, jun
  2002.

\bibitem{Dey2021}
Tamal~K Dey and Yusu Wang.
\newblock {\em Computational Topology for Data Analysis}.
\newblock Cambridge University Press, 2021.

\bibitem{Munch2017}
Elizabeth Munch.
\newblock A user's guide to topological data analysis.
\newblock {\em Journal of Learning Analytics}, 4(2):47--61, jul 2017.

\bibitem{Carlsson2010}
Gunnar Carlsson and Vin de~Silva.
\newblock Zigzag persistence.
\newblock {\em Foundations of Computational Mathematics}, 10(4):367--405, apr
  2010.

\bibitem{Aktas2019}
Mehmet~E. Aktas, Esra Akbas, and Ahmed~El Fatmaoui.
\newblock Persistence homology of networks: methods and applications.
\newblock {\em Applied Network Science}, 4(1), aug 2019.

\bibitem{Gallotti2015}
Riccardo Gallotti and Marc Barthelemy.
\newblock The multilayer temporal network of public transport in great britain.
\newblock {\em Scientific Data}, 2(1), jan 2015.

\bibitem{Tymochko2020}
Sarah Tymochko, Elizabeth Munch, and Firas~A. Khasawneh.
\newblock Using zigzag persistent homology to detect hopf bifurcations in
  dynamical systems.
\newblock {\em Algorithms}, 13(11):278, oct 2020.

\bibitem{McCullough2015}
Michael McCullough, Michael Small, Thomas Stemler, and Herbert Ho-Ching Iu.
\newblock Time lagged ordinal partition networks for capturing dynamics of
  continuous dynamical systems.
\newblock {\em Chaos: An Interdisciplinary Journal of Nonlinear Science},
  25(5):053101, may 2015.

\bibitem{Myers2019}
Audun Myers, Elizabeth Munch, and Firas~A. Khasawneh.
\newblock Persistent homology of complex networks for dynamic state detection.
\newblock {\em Physical Review E}, 100(2), aug 2019.

\bibitem{Myers2022}
Audun Myers, Firas~A. Khasawneh, and Elizabeth Munch.
\newblock Topological signal processing using the weighted ordinal partition
  network.
\newblock apr 2022.

\bibitem{Hatcher}
Allen Hatcher.
\newblock {\em Algebraic Topology}.
\newblock Cambridge University Press, 2002.

\bibitem{Munkres2}
James~R. Munkres.
\newblock {\em Elements of Algebraic Topology}.
\newblock Addison Wesley, 1993.

\bibitem{Oudot2015}
S.~Y. Oudot.
\newblock {\em Persistence theory: from quiver representations to data
  analysis}, volume 209 of {\em AMS Mathematical Surveys and Monographs}.
\newblock American Mathematical Society, 2015.

\bibitem{Carlsson2009a}
Gunnar Carlsson, Vin de~Silva, and Dmitriy Morozov.
\newblock Zigzag persistent homology and real-valued functions.
\newblock In {\em Proceedings of the 25th annual symposium on Computational
  geometry - SCG 09}. {ACM} Press, 2009.

\bibitem{Dionysus2}
Dmitriy Morozov.
\newblock Dionysus2.
\newblock \url{http://www.mrzv.org/software/dionysus2/}, 2019.

\bibitem{Landherr2010}
Andrea Landherr, Bettina Friedl, and Julia Heidemann.
\newblock A critical review of centrality measures in social networks.
\newblock {\em Business {\&} Information Systems Engineering}, 2(6):371--385,
  oct 2010.

\bibitem{vanHagen2011}
M.~{van Hagen}.
\newblock {\em Waiting experience at train stations}.
\newblock PhD thesis, University of Twente, April 2011.

\bibitem{Khor2016}
Alexander Khor and Michael Small.
\newblock Examining k-nearest neighbour networks: Superfamily phenomena and
  inversion.
\newblock {\em Chaos: An Interdisciplinary Journal of Nonlinear Science},
  26(4):043101, apr 2016.

\bibitem{Jacob2019}
Rinku Jacob, K.~P. Harikrishnan, R.~Misra, and G.~Ambika.
\newblock Weighted recurrence networks for the analysis of time-series data.
\newblock {\em Proceedings of the Royal Society A: Mathematical, Physical and
  Engineering Sciences}, 475(2221):20180256, jan 2019.

\bibitem{Small2009a}
Michael Small, Jie Zhang, and Xiaoke Xu.
\newblock Transforming time series into complex networks.
\newblock In {\em Lecture Notes of the Institute for Computer Sciences, Social
  Informatics and Telecommunications Engineering}, pages 2078--2089. Springer
  Berlin Heidelberg, 2009.

\bibitem{Small2013a}
Michael Small.
\newblock Complex networks from time series: Capturing dynamics.
\newblock In {\em 2013 {IEEE} International Symposium on Circuits and Systems
  ({ISCAS}2013)}. {IEEE}, may 2013.

\bibitem{Pomeau1980}
Yves Pomeau and Paul Manneville.
\newblock Intermittent transition to turbulence in dissipative dynamical
  systems.
\newblock {\em Communications in Mathematical Physics}, 74(2):189--197, jun
  1980.

\bibitem{Myers2020c}
Audun Myers and Firas~A. Khasawneh.
\newblock On the automatic parameter selection for permutation entropy.
\newblock {\em Chaos: An Interdisciplinary Journal of Nonlinear Science},
  30(3):033130, mar 2020.

\bibitem{CohenSteiner2006}
David Cohen-Steiner, Herbert Edelsbrunner, and Dmitriy Morozov.
\newblock {Vines and vineyards by updating persistence in linear time}.
\newblock {\em Proceedings of the twenty-second annual symposium on
  Computational geometry - SCG '06}, page 119, 2006.

\bibitem{Ulmer2019}
M.~Ulmer, Lori Ziegelmeier, and Chad~M. Topaz.
\newblock A topological approach to selecting models of biological experiments.
\newblock {\em {PLOS} {ONE}}, 14(3):e0213679, mar 2019.

\bibitem{Guezel2022}
{\.{I}}smail Güzel, Elizabeth Munch, and Firas~A. Khasawneh.
\newblock Detecting bifurcations in dynamical systems with {CROCKER} plots.
\newblock {\em Chaos: An Interdisciplinary Journal of Nonlinear Science},
  32(9):093111, sep 2022.

\bibitem{Kim2022}
Woojin Kim and Facundo Mémoli.
\newblock Extracting persistent clusters in dynamic data via m\"obius
  inversion, 2022.

\bibitem{Kim2019}
Woojin Kim, Facundo Mémoli, and Anastasios Stefanou.
\newblock Interleaving by parts: Join decompositions of interleavings and
  join-assemblage of geodesics.
\newblock December 2019.

\bibitem{Kim2020}
Woojin Kim, Facundo M{\'e}moli, and Zane Smith.
\newblock Analysis of dynamic graphs and dynamic metric spaces via zigzag
  persistence.
\newblock In Nils~A. Baas, Gunnar~E. Carlsson, Gereon Quick, Markus Szymik, and
  Marius Thaule, editors, {\em Topological Data Analysis}, pages 371--389,
  Cham, 2020. Springer International Publishing.

\bibitem{Kim2020a}
Woojin Kim.
\newblock {\em The Persistent Topology of Dynamic Data}.
\newblock phdthesis, The Ohio State University, 2020.

\bibitem{Kim2021}
Woojin Kim and Facundo M{\'e}moli.
\newblock Spatiotemporal persistent homology for dynamic metric spaces.
\newblock {\em Discrete \& Computational Geometry}, 66(3):831--875, 2021.

\bibitem{Chowdhury2018c}
Samir Chowdhury, Bowen Dai, and Facundo Mémoli.
\newblock The importance of forgetting: Limiting memory improves recovery of
  topological characteristics from neural data.
\newblock {\em PLOS ONE}, 13(9):1--20, 09 2018.

\bibitem{Milosavljevic2011}
Nikola Milosavljevic, Dmitriy Morozov, and Primoz Skraba.
\newblock Zigzag persistent homology in matrix multiplication time.
\newblock In {\em Proceedings of the 27th Annual Symposium on Computational
  Geometry}, 2011.

\bibitem{Bauer2021}
Ulrich Bauer.
\newblock Ripser: efficient computation of vietoris{\textendash}rips
  persistence barcodes.
\newblock {\em Journal of Applied and Computational Topology}, jun 2021.

\bibitem{Otter2017}
Nina Otter, Mason~A Porter, Ulrike Tillmann, Peter Grindrod, and Heather~A
  Harrington.
\newblock A roadmap for the computation of persistent homology.
\newblock {\em {EPJ} Data Science}, 6(1), aug 2017.

\bibitem{Dey2021a}
Tamal~K. Dey and Tao Hou.
\newblock Computing zigzag persistence on graphs in near-linear time.
\newblock March 2021.

\bibitem{Dey2022}
Tamal~K. Dey and Tao Hou.
\newblock Fast computation of zigzag persistence.
\newblock April 2022.

\bibitem{Dey2021b}
Tamal~K. Dey and Tao Hou.
\newblock Updating barcodes and representatives for zigzag persistence.
\newblock December 2021.

\end{thebibliography}
